\documentclass[acmsmall,screen]{acmart}
\AtBeginDocument{%
  }

\setcopyright{acmlicensed}
\copyrightyear{2026}
\acmYear{2026}
\acmDOI{XXXXXX}

\acmJournal{TACO}
\acmVolume{1}
\acmNumber{1}
\acmArticle{1}
\acmMonth{1}

\usepackage{microtype}
\usepackage{graphicx}
\usepackage{caption}
\usepackage{subcaption}
\usepackage{booktabs} 
\usepackage{xspace}
\usepackage{multirow}
\usepackage[table]{xcolor}
\usepackage{enumitem}

\usepackage{pifont}
\usepackage{hyperref}

\usepackage[normalem]{ulem}

\usepackage{algorithm}
\usepackage{algorithmic}

\usepackage{amsmath}
\usepackage{mathtools}
\usepackage{amsthm}
\usepackage{amsfonts}

\usepackage[capitalize,noabbrev]{cleveref}

\theoremstyle{plain}

\theoremstyle{definition}

\theoremstyle{remark}

\newcommand{\sys}{CoCoScale}

\renewcommand\footnotetextcopyrightpermission[1]{}

\begin{document}

\title{\sys{}: Leveraging Layer-wise Scaling to Unlock the Potential of Online LLM Serving}

\author{Jingfeng Wu}
\authornote{Both authors contributed equally to this research.}
\email{jf.wu2@siat.ac.cn}
\author{Yiyuan He}
\authornotemark[1]
\email{yy.he2@siat.ac.cn}

\author{Minxian Xu}
\email{mx.xu@siat.ac.cn}
\authornote{Minxian Xu is the corresponding author.}
\author{Xitong Gao}
\email{xt.gao@siat.ac.cn}
\affiliation{%
  \institution{Shenzhen Institutes of Advanced Technology, Chinese Academy of Sciences}
  \city{Shenzhen}
  \country{China}
}

\author{Chong Ma}
\email{machong.mc@alibaba-inc.com}
\author{Le Chen}
\email{donghuai.cl@taobao.com}
\author{Min Shen}
\email{shenmin.sm@taobao.com}
\author{Lin Qu}
\email{xide.ql@taobao.com}
\affiliation{%
  \institution{Alibaba Group Inc}
  \city{Hangzhou}
  \country{China}
}

\author{Kejiang Ye}
\email{kj.ye@siat.ac.cn}
\affiliation{%
  \institution{Shenzhen Institutes of Advanced Technology, Chinese Academy of Sciences}
  \city{Shenzhen}
  \country{China}
}
  
\author{Chengzhong Xu}
\email{czxu@um.edu.mo}
\affiliation{%
  \institution{State Key Lab of IOTSC, University of Macau}
  \city{Macau SAR}
  \country{China}
}

\renewcommand{\shortauthors}{Wu et al.}


\begin{abstract}
Online large language model (LLM) serving has become the backbone of modern AI applications, powering diverse downstream services through shared hardware clusters. However, modern serving systems frequently encounter highly dynamic workloads characterized by severe workload skewness, where a small fraction of model instances receives the vast majority of traffic. Existing instance-level scaling mechanisms are limited by coarse-grained resource adjustment: scaling up requires the cold-start of full-model replicas, incurring substantial latency, while scaling down leaves the system vulnerable to performance degradation during sudden traffic surges. The key insight of this work is that LLM serving offers a unique opportunity for fine-grained scaling. In this paper, we propose CoCoScale, a layer-wise dynamic scaling mechanism that selectively expands the parallelism of hot layers onto idle resources reclaimed from underutilized devices, enabling elastic data parallelism without altering model architectures or adding hardware overhead. Evaluations demonstrate that CoCoScale significantly reduces cold start latency by 97.9\%--99.3\% compared to traditional scale up. Under production traces, CoCoScale reduces average latency by 20.7\%--28.1\% and achieves full Service Level Objective (SLO) attainment, demonstrating superior dynamic adaptability and resource efficiency. 
\end{abstract}



\maketitle

\section{Introduction}

The advent of state-of-the-art LLMs, such as GPT 5.2~\cite{gpt5.2}, Qwen 3~\cite{qwen}, and DeepSeek V3~\cite{DeepSeekV3TR}, has reshaped user experiences and driven the rapid growth of online LLM serving. Consequently, major cloud platforms~\cite{Ali-cloud,AzureML,VertexAI} have emerged as the foundational infrastructure for diverse downstream applications.

Despite these advancements, online serving environments contend with significant operational challenges stemming from extreme workload volatility and imbalance. Specifically, request patterns exhibit unpredictable dynamics, leading to frequent load fluctuations across instances. Furthermore, there is a severe workload disparity among heterogeneous models. As demonstrated by the OpenRouter~\cite{OpenRouter} leaderboard, the top 25\% of model instances process 75\% of the total traffic, creating a distinct ``hot/cold skewness.'' This imbalance arises from entrenched user preferences for specific models, making it difficult to alleviate through simple request migration on the provider side without violating user-model affinity.

\begin{figure}[!ht]
    \centering
    \includegraphics[width=0.6\linewidth]{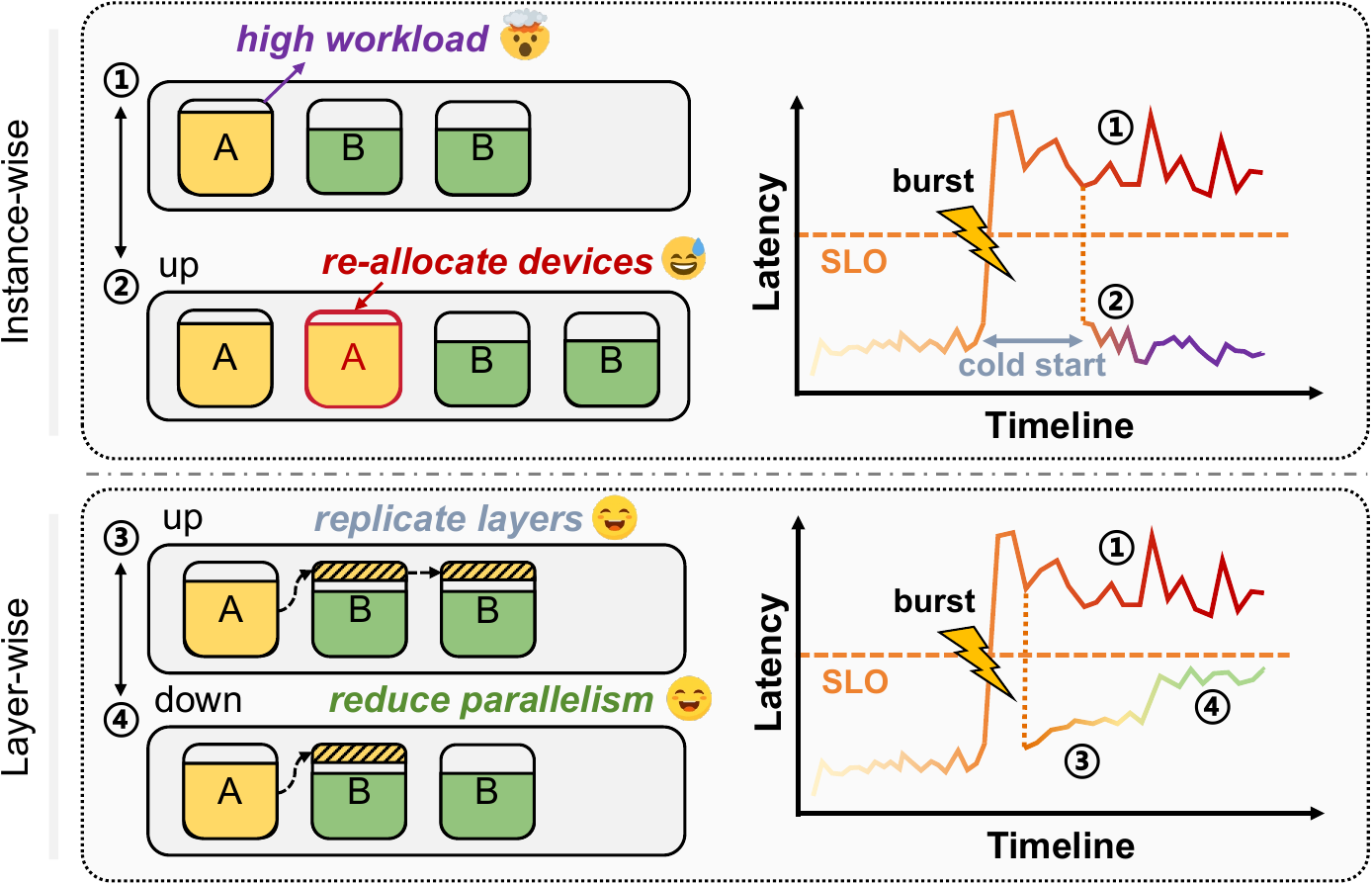}
    \caption{Comparison between instance-wise and layer-wise scaling. (1) Instance-level scaling suffers from significant latency spikes due to cold starts. (2) CoCoScale enables rapid SLO recovery by selectively replicating layers onto idle resources.}
    \label{fig:intro}
\end{figure}

Mitigating these challenges requires robust runtime scaling mechanisms capable of dynamically reconfiguring resource allocation. However, existing instance-wise scaling strategies are fundamentally restricted by their coarse granularity. As illustrated in Figure \ref{fig:intro}, addressing a traffic burst requires the re-allocation of full model replicas, as shown in \ding{173}. This rigid process triggers a prohibitive cold-start period for container initialization and parameter loading, resulting in provisioning latency that significantly lags behind demand spikes and leads to temporary SLO violations~\cite{BlitzScale}. Additionally, since resources can only be scaled in large and fixed units, such as entire instances, it is impossible to perfectly match the fluctuating workload. This lack of precision forces the system to over-provision resources, leaving valuable computing power locked in underutilized instances. In contrast, our proposed approach, CoCoScale,  introduces layer-wise scaling to eliminate these overheads. By selectively expanding the parallelism of hot layers onto reclaimed idle resources, as depicted in states \ding{174} and \ding{175}, our approach delivers near-instant performance boosts. This fine-grained elasticity enables the system to strictly adhere to SLOs while minimizing resource waste, bypassing the operational delays inherent in traditional methods.

To overcome these limitations, we propose layer-wise scaling, a fine-grained mechanism that operates without triggering costly full-model redeployments. The key insight is to reclaim idle resource from ``cold" devices to boost the performance of ``hot" instances. This achieves a dual effect: scaling up the throughput of hot instances while effectively consolidating fragmented idle resources on cold ones.

In particular, we replicate selected layers of a hot instance onto underutilized devices. These replicas perform inference alongside the original layers, forming layer-wise data parallelism. This approach allows the system to dynamically adjust the parallelism degree of specific layers in response to workload fluctuations, thereby satisfying SLO requirements with precision. Crucially, by leveraging communication-computation overlapping, layer-wise scaling maintains a minimal communication footprint, ensuring that the performance gains are not eroded by synchronization overhead. Thanks to its fine-grained granularity, layer-wise scaling incurs significantly lower setup latency compared to instance-level scaling, making it a robust alternative when coarse-grained methods fall short.

In this paper, we present \sys{}, the layer-wise scaling mechanism and its corresponding serving framework. By replicating specific layers onto reclaimed idle resources, \sys{} enables fine-grained elastic data parallelism that eliminates the need for full-model redeployment. We implemented a prototype of \sys{} based on the Nano-vLLM~\cite{nano-vllm} backend and conducted extensive evaluations using production-scale traces. The results demonstrate that \sys{} achieves a 97.9\%-99.3\% reduction in scaling latency and ensures 100\% SLO attainment. Furthermore, the system reduces average end-to-end (E2E) latency by 20.7\%--28.1\% over baseline methods.

In summary, our primary contributions are as follows:
\begin{itemize}[noitemsep, topsep=0pt, parsep=0pt, partopsep=0pt]
\item We conduct a systematic characterization of workload challenges in online LLM serving, identifying the staircase lag and quantization mismatch inherent in coarse-grained scaling.
\item We propose layer-wise scaling mechanism and implement a system framework that decouples performance adjustment from rigid instance boundaries to enable rapid, non-integer capacity expansion.
\item Extensive evaluations validate that \sys{} significantly enhances system efficiency while maintaining stable performance under varying workloads.
\end{itemize}

The rest of the paper is organized as follows: Section \ref{sec:bg} discusses the background and motivation of the proposed approach, and the mechanism that supports layer-wise replication is introduced in Section \ref{sec:replication_sec}. Section \ref{sec:modeling_sec} presents the layer-wise algorithm for scaling resources and Section \ref{sec:sys} demonstrates the system design of \sys{}. Performance evaluations are illustrated in Section \ref{sec:evaluation}. The  related work are summarized in Section \ref{sec:related}, and Section 8 concludes the paper.

\section{Background and Motivation}
\label{sec:bg}
\subsection{Online LLM Serving}
Online LLM serving provides a scalable infrastructure that allows users to integrate advanced intelligence into applications through simple APIs without managing underlying hardware \cite{xu2026cloudnativedistributedsystemsefficient}. The core of modern online serving is LLM inference, which requires providing high-concurrency and highly reliable services for users on shared hardware clusters. This presents significant challenges to the entire architecture. On one hand, the system must efficiently manage computing and storage resources to avoid workload imbalance. On the other hand, the system also needs to meet strict SLOs for different businesses while balancing high throughput and minimized end to end latency.

\subsection{Workload Skewness and Variety}


The workloads in online LLM serving exhibit a high degree of hot and cold skewness as well as burstiness. 
Since model instances with different popularity are heterogeneous, traditional schedulers~\cite{osdi24-llumnix,icsoc24-uellm} cannot direct requests from hot instances to cold ones. This situation eventually leads to severe load imbalance among the hardware devices to which these instances belong.
In addition, the request patterns for the same model show a wide range of variability and are difficult to predict. From a macro perspective, instances face periodic changes in load levels, such as the tidal phenomenon. From a micro perspective, instances occasionally encounter sudden bursts of traffic, which places high demands on the elastic scaling capabilities of the system.

\subsection{Limitations of Instance-wise Scaling}

\begin{figure}[h]
    \centering
    \includegraphics[width=\linewidth]{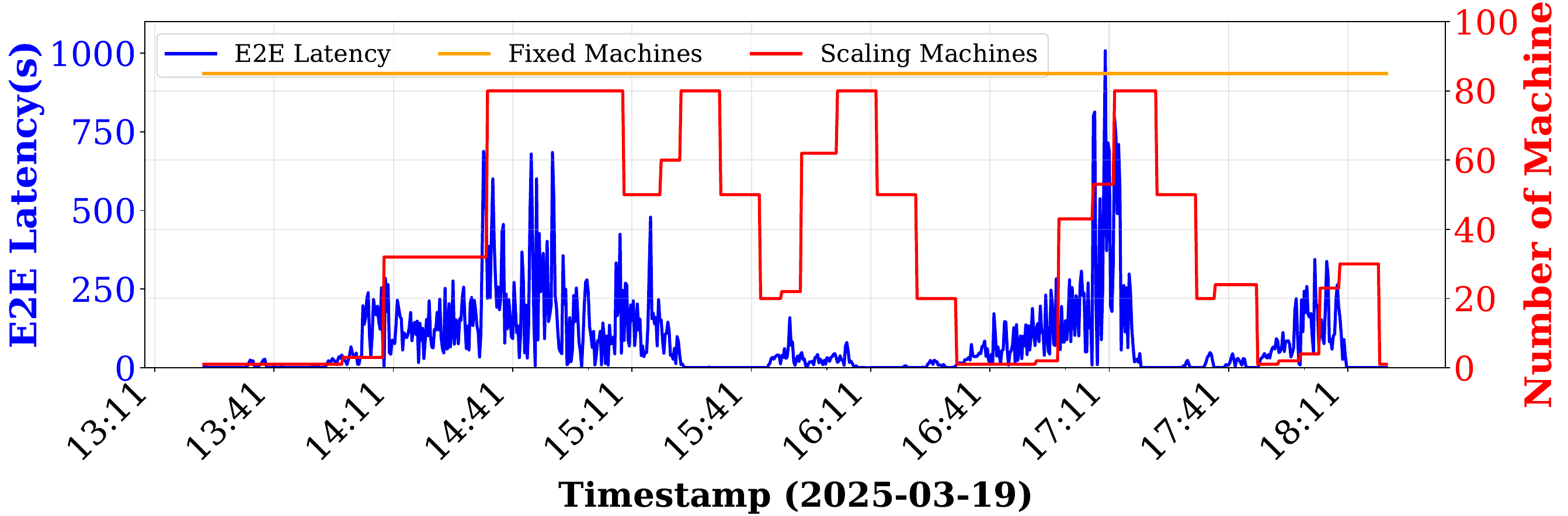}
    \caption{Production traces highlighting the \textbf{staircase pattern} and \textbf{temporal lag} in coarse-grained scaling relative to request spikes.}
    \Description{A plot showing the staircase pattern and temporal lag in scaling devices compared to request spikes.}
    \label{fig:coarse_scaling}
\end{figure}

To address these challenges, contemporary LLM serving systems rely on elastic scaling at the instance level~\cite{zeng2025medusa,yang2022infless,lin2025flexpipe}, where entire model replicas are deployed or removed across GPUs to accommodate workload changes. However, as shown in the production traces from Alibaba in Figure \ref{fig:coarse_scaling}, these coarse-grained strategies face significant hurdles~\cite{BlitzScale,ServerlessLLM}. A primary obstacle is the high provisioning latency caused by the inherent cold-start problem. Because initializing a new model instance involves time-consuming operations like container setup and loading massive weights, the system exhibits a pronounced temporal lag. This is visible in Figure \ref{fig:coarse_scaling}, where the step-like progression of Scaling Devices fails to keep pace with sharp spikes in response time, resulting in frequent SLO violations.

Beyond latency, existing methods also suffer from a lack of precision that leads to resource inefficiency. Since scaling occurs in large increments such as adding a complete GPU or model instance, the system often struggles to match SLOs exactly. This lack of granularity results in over-provisioning where added capacity exceeds actual demand. While aggressive scaling might eventually stabilize response times, the gains are negated by excessive resource consumption because current architectures cannot finely tune performance to follow the exact demand curve. The fundamental issue is the mismatch between the massive, indivisible model instance and the highly dynamic, fine-grained workload fluctuations. This mismatch suggests that breaking down the scaling unit from the entire model to its internal modular layers could provide the necessary agility and precision for modern LLM serving.

These observations motivate a shift in perspective: instead of treating the entire model instance as the atomic unit of scaling, we propose to operate at the granularity of individual Transformer layers. In the following sections, we introduce layer-wise replication as a mechanism for fine-grained data parallelism, present a formal modeling framework to characterize its performance trade-offs, and describe a complete system design that realizes this vision in production environments.
\section{Layer-wise Replication}\label{sec:replication_sec}

Inspired by the concept of instance-level data parallelism (DP), we explore its application at a finer granularity. Specifically, we adopt layer-wise replication as a fine-grained augmentation for Transformer layers (also referred to as blocks) to implement layer-wise DP~\cite{osdi22-alpa,osdi23-alpaserve}. Unlike instance-level scaling, which treats the entire model as an indivisible unit, layer-wise replication selectively duplicates a subset of layers onto available devices. This section describes the replication mechanism and the optimized data transfer protocols that make it practical.

\subsection{Replication Mechanism}\label{sec:replication}

\begin{figure}[htbp]
    \centering
    \includegraphics[width=0.9\linewidth]{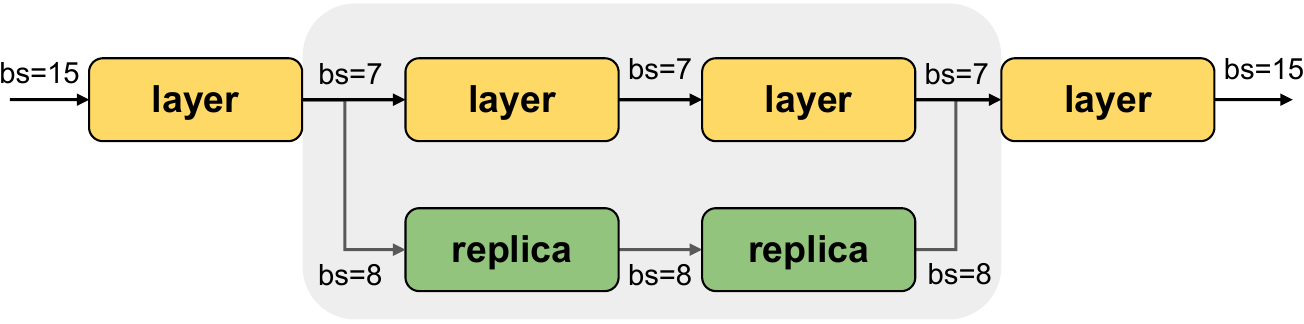}
    \caption{Illustration of the replication for decoder layers. Yellow blocks represent the layers deployed on the main device, while green blocks represent replicas deployed on another device.}
    \label{fig:replication}
\end{figure}

As illustrated in Figure \ref{fig:replication}, by replicating the weights and KV caches of selected layers onto available devices, the system creates layer replicas. This enables a more precise workload distribution where a batch size (bs) of 15 requests is split into sub-batches of 7 and 8. These sub-batches are processed concurrently by the original layers and their replicas, with results merged after completing the parallel blocks. This approach establishes the foundation for layer-wise DP, where the resulting speedup and communication overhead are determined by the number of involved layers ($N$) and the degree of parallelism ($P$).

To minimize runtime communication overhead, we treat multiple consecutive layers as a monolithic unit for replication. As illustrated by the gray shaded area in Figure \ref{fig:replication}, this coarse-grained strategy ensures that intermediate activations generated by replicas remain resident in local memory. Consequently, cross-device transfers are confined exclusively to the segment boundaries, making the transmission overhead fixed relative to the replicated layer count $N$. Given this design of mutually independent replicas, we are able to enable the static graph execution mode for these replicas, as described in Section \ref{sec:compat}.

At the input boundary, the host device scatters the incoming batch into sub-batches according to the parallelism degree $P$. Each replica independently processes its assigned sub-batch through the $N$ consecutive layers without any inter-device synchronization. At the output boundary, the partial results from all replicas are gathered back to the host device and concatenated along the batch dimension to reconstruct the full output tensor. This scatter-gather pattern requires exactly two communication rounds per forward pass, regardless of the number of replicated layers, which keeps the communication overhead predictable and bounded.

For the KV cache, each replica maintains its own partition corresponding to the sub-batch it processes. During autoregressive decoding, new key-value entries are appended locally on each device without cross-device coherence traffic. This partitioned design avoids the synchronization overhead that would arise from maintaining a globally consistent cache. When the parallelism configuration changes (e.g., increasing $P$ from 2 to 4), the existing cache entries must be redistributed to match the new sub-batch assignment. This redistribution is performed as part of the configuration transition protocol, leveraging the same chunked transfer mechanism described in Section \ref{sec:transfer} to overlap cache migration with ongoing decoding steps.

\subsection{Optimized Data Transfer}\label{sec:transfer}

\begin{figure*}[htbp]
    \centering
    \includegraphics[width=\linewidth]{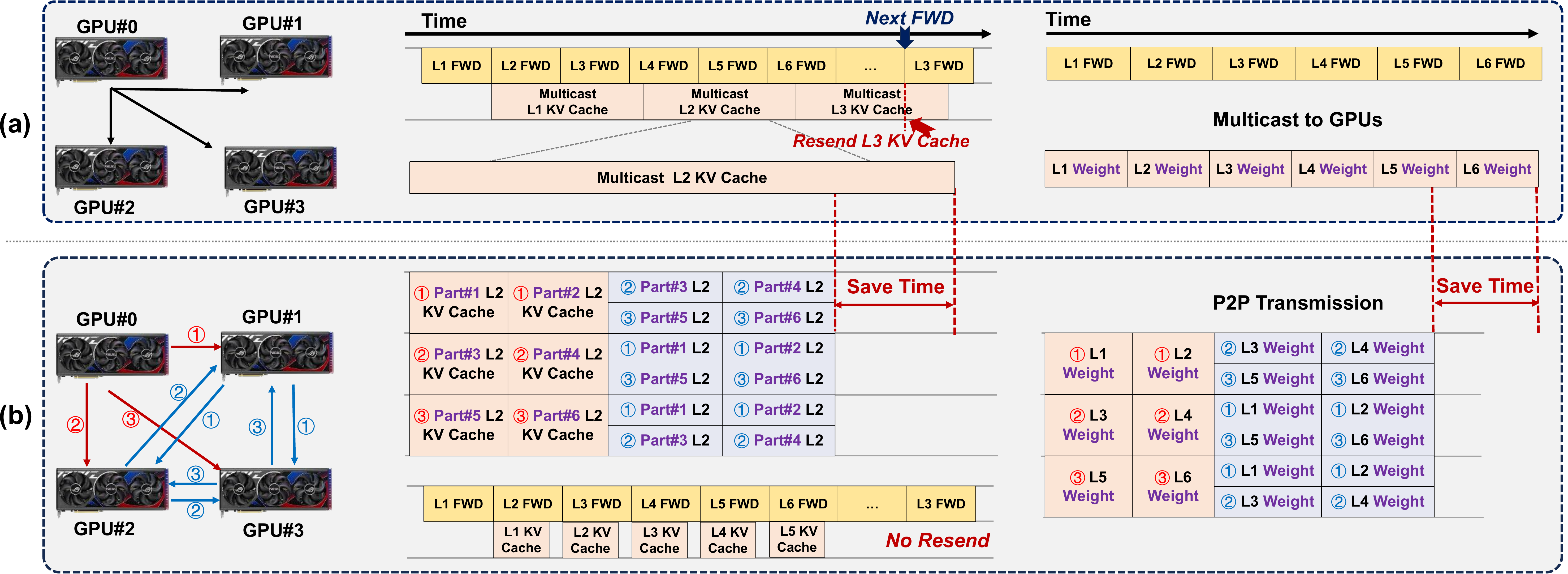}
    \caption{Comparison of data transfer strategies. (a) Single-source multicast causes linear delay and ``Resend'' conflicts. (b) CoCoScale utilizes a \textbf{Scatter-then-Exchange} ring topology. Numbered arrows denote concurrent transmission phases of different chunks utilizing full-duplex bandwidth.}
    \label{fig:transfer}
\end{figure*}

To minimize the setup latency during replica creation, it is essential to maximize the utilization of P2P network bandwidth when transferring weights and the KV Cache. As shown in Figure \ref{fig:transfer}(a), in the absence of full duplex communication (e.g., NVLink), standard data transfer schemes typically employ a single source multicast approach. This sequential execution creates a significant performance bottleneck where the total transmission time grows linearly with the number of GPUs. Furthermore, the prolonged transmission of dynamic KV Cache data often overlaps with the subsequent forward pass. This conflict forces the system to abort and resend incomplete data, resulting in unpredictable scaling delays.

To maximize bandwidth utilization, we design a bidirectional chunked multicast mechanism based on a ring topology~\cite{dean2008mapreduce,Patel2023SplitwiseEG}. As illustrated in the weight transfer flow of Figure \ref{fig:transfer}(b), we employ a Scatter-then-Exchange strategy. The source GPU first partitions the weights and scatters distinct segments to different neighbors simultaneously. These segments then circulate across the ring in both clockwise and counter-clockwise directions, leveraging the full-duplex capabilities of NVLink to significantly reduce the broadcasting latency compared to sequential unicast.

For dynamic KV Cache states, we adopt a fine-grained chunked transfer strategy to ensure timing precision. As depicted by the blocks (e.g., Part\#1 through Part\#6) in Figure \ref{fig:transfer}(b), the KV cache is partitioned and transmitted via parallel streams. This fine granularity allows data transfer to fit strictly within the narrow inter-token intervals. By masking communication behind computation, this design effectively prevents the temporal overlap conflicts with the next forward pass, eliminating the need for costly data resending.
\section{Modeling and Scaling Algorithm}\label{sec:modeling_sec}

The acceleration capability of layer-wise DP constitutes the core of our fine-grained scaling mechanism. In this section, we first conduct systematic ablation studies to evaluate the sensitivity of inference performance to varying layer counts ($N$) and parallelism degrees ($P$). Guided by these empirical insights, we then establish a formal modeling framework that captures the speedup and communication overhead of layer-wise DP. Finally, we present a unified scaling algorithm that leverages this analytical model to determine optimal configurations in real time.

\subsection{Ablation Study of Layer-wise DP}\label{sec:ablation}

To analyze the performance characteristics and overhead of layer-wise DP, we denote the number of consecutive layers replicated as $N$, and the degree of parallelism as $P$, which together define a non-integer speedup ratio. For instance, in a 60-layer Qwen-32B model, a configuration of $N=20$ and $P=2$ can theoretically achieve a speedup approaching 120\%, offering a significant performance boost even when full instance-level scaling is infeasible.

To explore the marginal benefits of $N$ and $P$, we conducted ablation studies under various Requests Per Second (RPS) on four NVIDIA H20 GPUs using the Qwen-32B model. We investigated the impact of varying $N$ while maintaining a constant $P$. We used vLLM~\cite{sosp23-vllm} as the baseline and evaluated performance on the Alpaca dataset~\cite{23-alpaca} (with a fixed input length of 1k tokens and output length of 64 tokens), adopting end-to-end latency as the primary performance metric.

\begin{figure}[h]
    \centering
    \begin{subfigure}[b]{0.48\linewidth}
        \centering
        \includegraphics[width=\linewidth]{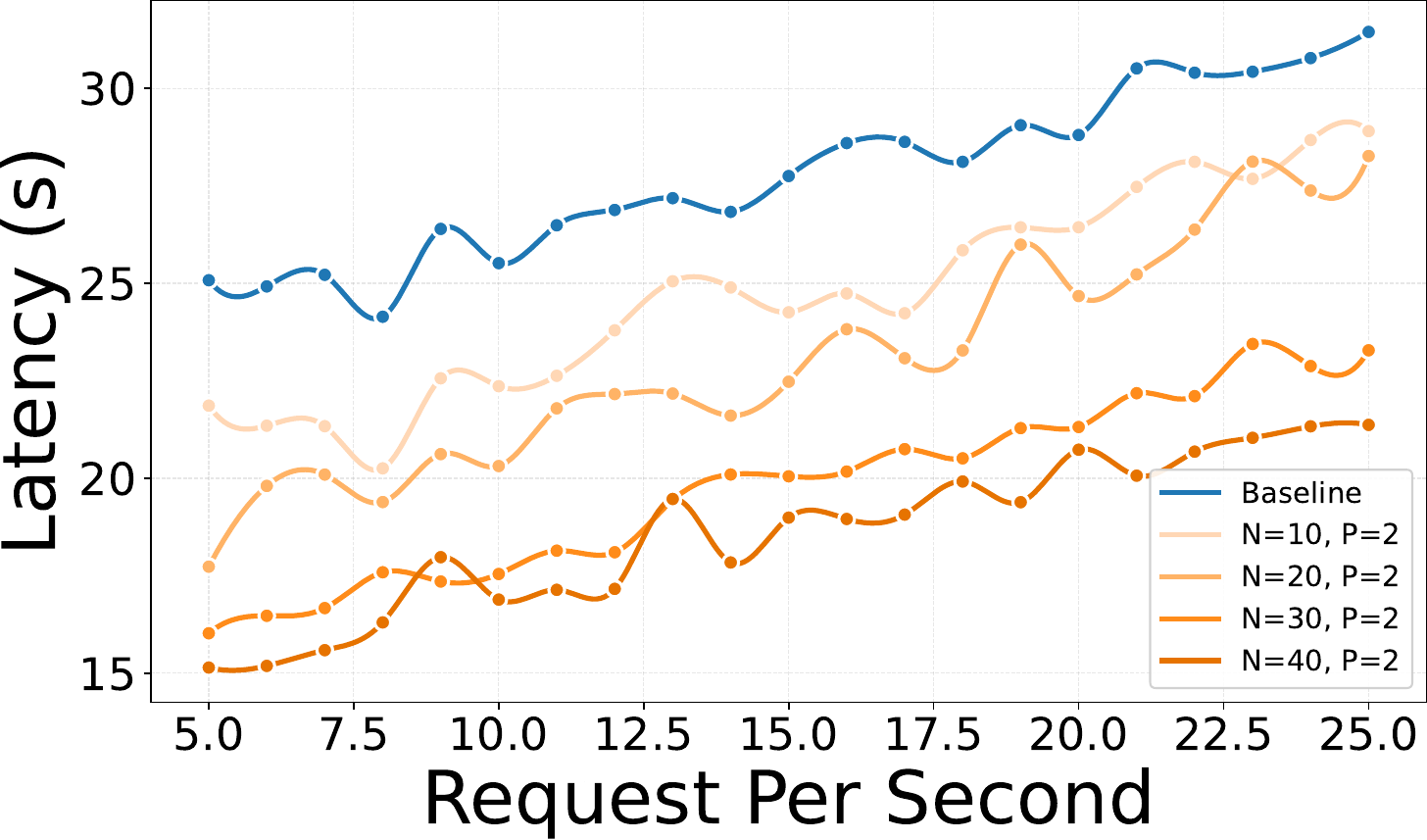}
        \caption{End-to-end Latency with varying $N$ (P=2)}
        \label{fig:layer_rep_throughput}
    \end{subfigure}
    \hfill
    \begin{subfigure}[b]{0.48\linewidth}
        \centering
        \includegraphics[width=\linewidth]{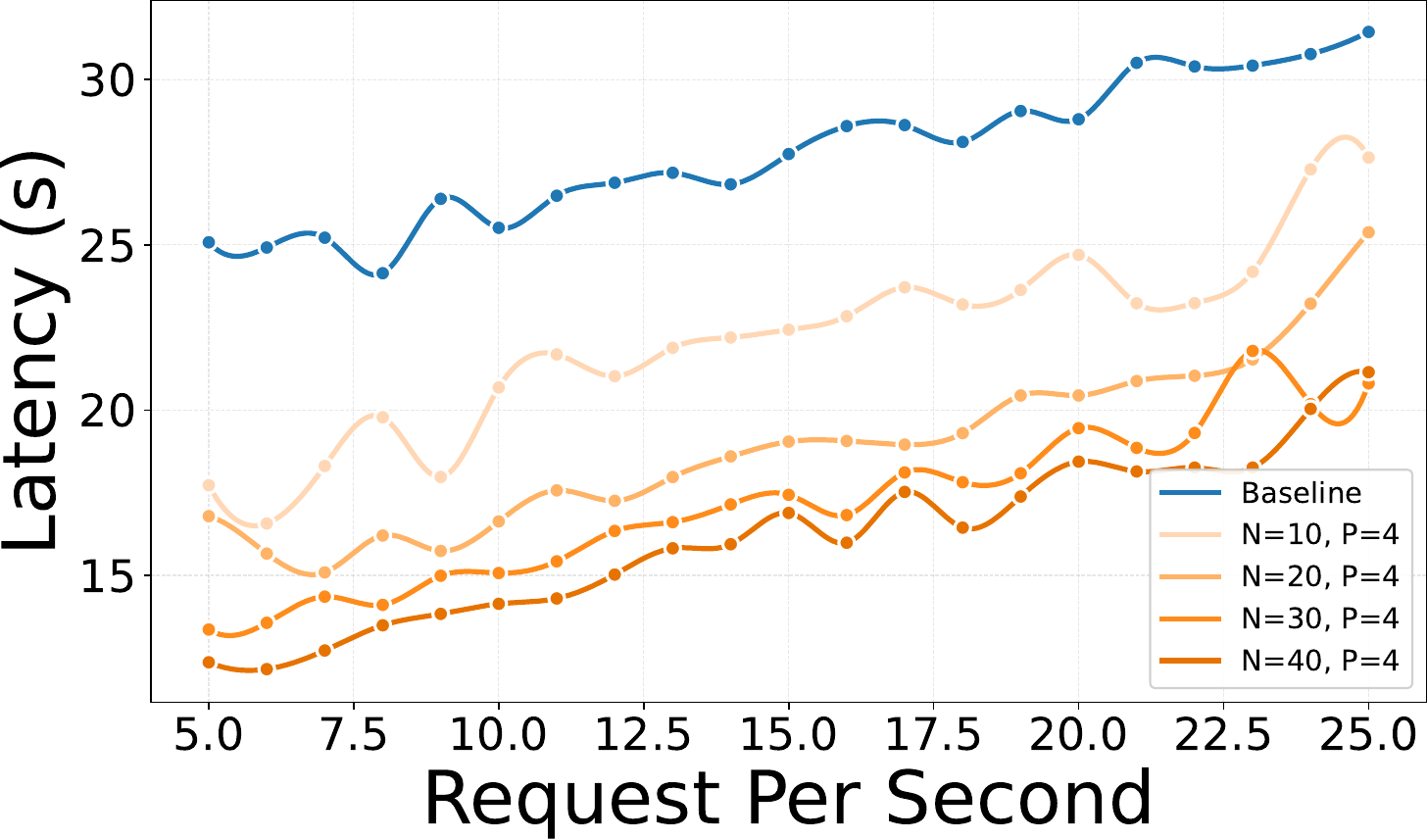}
        \caption{End-to-end Latency with varying $N$ (P=4)}
        \label{fig:layer_rep_latency}
    \end{subfigure}

    \caption{Performance analysis of different number of continuous replicas and parallelism under varying request rates.}
    \label{fig:replication_analysis}
\end{figure}

\textbf{Effect of Layer Count $N$.} With parallelism fixed at $P=2$, as $N$ increases from 10 to 40, the minimum improvement rises from 4.3\% to 12.4\%, and the maximum improvement from 15.4\% to 38.7\%. Similarly, at $P=4$, the minimum improvement increases from 11.3\% to 32.7\%, while the maximum grows from 33.51\% to 51.2\%. This indicates that at a fixed parallelism level, increasing $N$ to cover more Transformer layers significantly reduces computational latency. However, as $N$ continues to grow, the gap between performance curves narrows, exhibiting a phenomenon of diminishing marginal returns. This suggests that the residual components, specifically those that cannot be parallelized or require redundant computation, gradually emerge as bottlenecks, limiting further performance gains.

\textbf{Effect of Parallelism $P$.} Increasing $P$ enhances the system's throughput under high concurrency but introduces substantial cross-device communication overhead. For example, with $N=20$, the theoretical maximum speedups for $P=2$ and $P=4$ are 33\% and 100\%, respectively. However, empirical results show maximum gains of only 29.3\% and 40.4\%. This discrepancy highlights that as $P$ increases, the latency overhead from device synchronization and data exchange constitutes a growing proportion of total execution time, thereby diluting the benefits of fine-grained data parallelism.

\begin{figure}[h]
    \centering
    \begin{subfigure}[b]{0.48\linewidth}
        \centering
        \includegraphics[width=\linewidth]{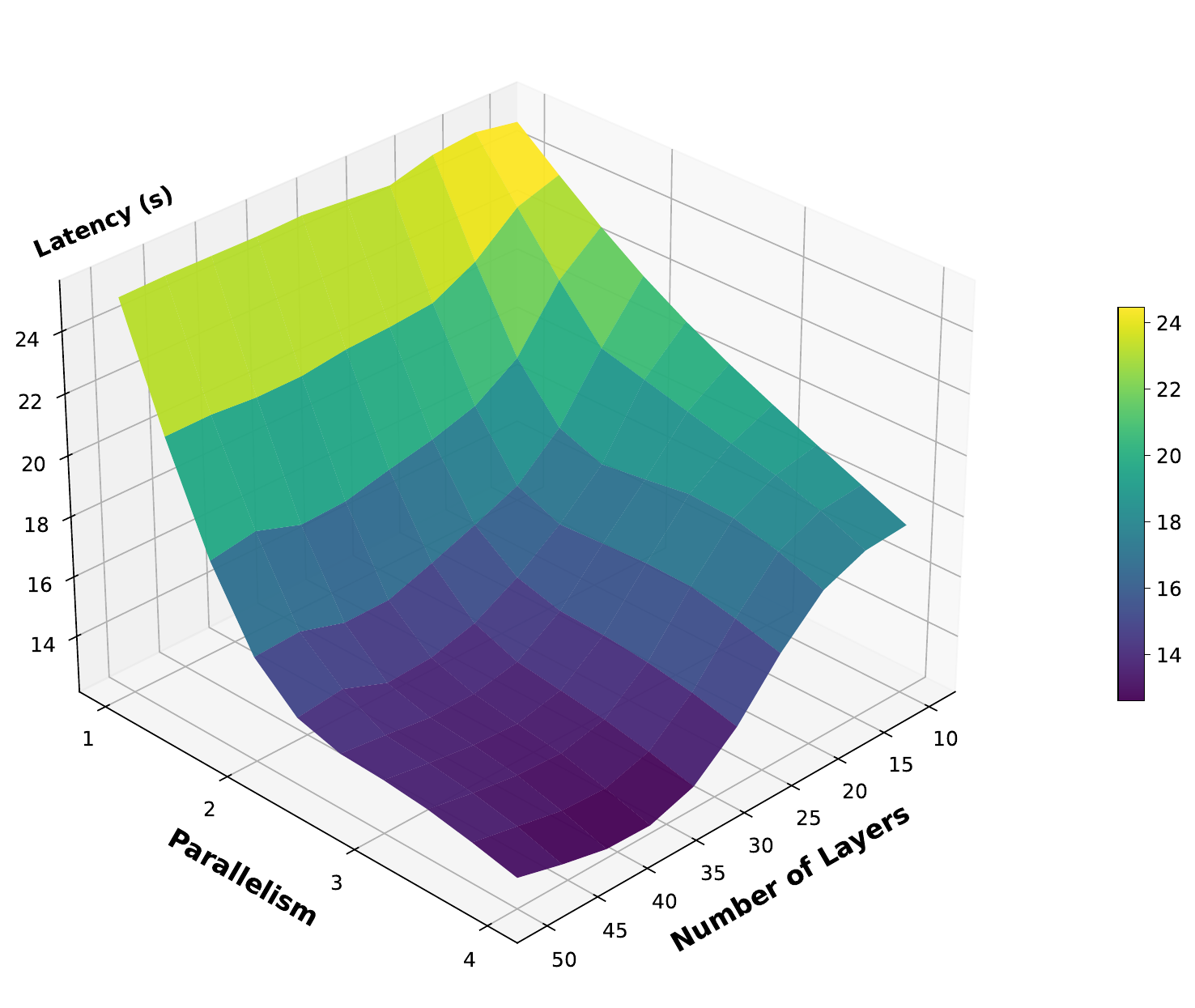}
        \caption{Low-concurrency (RPS=5)}
    \end{subfigure}
    \hfill
    \begin{subfigure}[b]{0.5\linewidth}
        \centering
        \includegraphics[width=\linewidth]{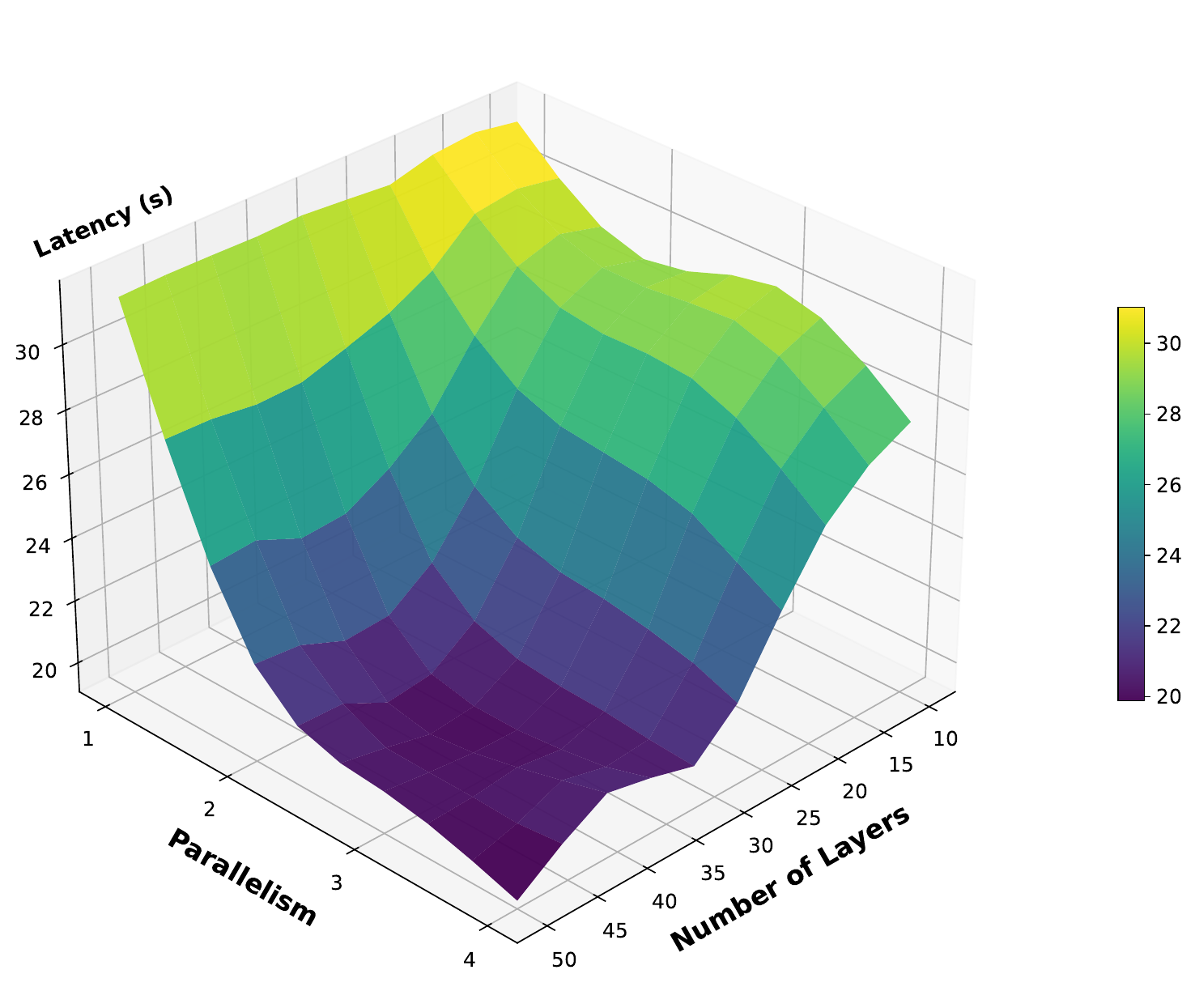}
        \caption{High-concurrency (RPS=25)}
    \end{subfigure}

    \caption{3D surface plots of end-to-end latency as a function of layer count $N$ and parallelism $P$ under two representative load levels.}
    \label{fig:3D}
\end{figure}

To visualize the joint effect of $N$ and $P$, Figure \ref{fig:3D} presents 3D surface plots of end-to-end latency under two representative load levels. At low concurrency (RPS=5), the latency surface is relatively flat: moderate values of $N$ and $P$ already bring the system close to its performance ceiling, and further increases yield marginal gains. In contrast, at high concurrency (RPS=25), the surface exhibits a pronounced valley where larger $N$ and moderate $P$ jointly reduce latency. Beyond this valley, however, increasing $P$ further provides diminishing returns as communication overhead grows. Notably, the optimal operating region shifts with load intensity, underscoring the need for an adaptive strategy that continuously adjusts both $N$ and $P$ in response to workload dynamics.

These empirical observations validate the effectiveness of layer-wise parallelism, serving as the cornerstone for our proposed scaling mechanism. Furthermore, compared to model-level parallelism, layer-wise parallelism introduces the number of layers as a new dimension. This flexibility offers a broader configuration space, liberating the system from rigid integer-level parallelism constraints. However, this increased flexibility also introduces complexity in decision-making. Consequently, we next introduce a formal model to analyze the speedup ratios and communication overheads across these diverse granularity configurations.

\subsection{Cost and Speedup Modeling}\label{sec:model}

To systematically generalize the empirical findings from our ablation studies and provide a rigorous basis for system optimization, this subsection establishes a formal modeling framework. We structure the analysis by decomposing the total latency into computation and communication components, followed by deriving the speedup ratio based on an adaptation of Amdahl's Law. While the classical formulation defines speedup as $S(a,p) = \frac{1}{(1-a)+\frac{a}{p}}$ (where $a$ represents the parallelizable fraction and $p$ denotes the speedup factor), our model extends this framework to quantify the specific trade-offs between the replicated layer count ($N$) and parallelism degree ($P$) in LLM serving.

\textbf{Assumptions and Symbol Definitions.}
To establish a formal framework, we define $L$ as the total number of layers in the model. A replica configuration is characterized by $N$, the number of consecutive layers selected for replication, and $P$, the degree of parallelism applied to these layers. Let $d_{hidden}$ denote the hidden dimension, $bs$ the batch size, and $B$ the interconnect bandwidth. We assume a homogeneous cluster environment where compute capacity $t_{comp}$ per layer and bandwidth $B$ are uniform across all devices. Additionally, we assume that sub batches are equally distributed among all replicas.

\textbf{Cost Modeling.}
The deployment of a configuration $(N, P)$ is physically constrained by the memory capacity and floating-point throughput of each participating device. It is important to clarify that a configuration of parallelism $P$ implies the cluster utilizes $P$ devices in total (1 host plus $P-1$ replicas) to process the $N$ layers. For modern LLMs employing SwiGLU structures (e.g., Qwen), the per-device resource consumption is modeled as follows. The value of $l$ varies between inference phases, corresponding to the prompt length during the prefill stage and a single token ($l=1$) during the decoding stage~\cite{hu2024inferenceinterferencedisaggregatellm,qin2024mooncakekvcachecentricdisaggregatedarchitecture}. Let $\kappa$ denote the parameter storage coefficient (typically 2 bytes for FP16). The memory cost $C_{mem}$ and computational load $C_{comp}$ required for a single device are approximated as:
\begin{equation}
C_{mem}(N,P) \approx N \cdot \kappa\cdot(12 d_{hidden}^2 + \frac{bs}{P} \cdot l \cdot d_{hidden}),
\label{eq:cost_mem}
\end{equation}
\begin{equation}
C_{comp}(N,P) \approx \frac{N}{P} \cdot 24 \cdot bs \cdot l \cdot d_{hidden}^2.
\label{eq:cost_com}
\end{equation}
This formulation reflects the mechanics of layer-wise replication: while the static weight memory per device remains constant (each replica holds a full copy), the dynamic activation memory and computational load are amortized by the total parallelism degree $P$.

\textbf{Speedup Modeling.} Based on the resource overhead, we quantify the trade-offs between the replicated layer count $N$ and parallelism $P$. Let $L$ be the total number of layers and $t_{comp}$ the computation time per layer. The total inference latency is defined as the sum of the computation cost $W(N, P)$ and the communication overhead $T(P)$:\begin{equation}\text{Latency}(N, P) = W(N, P) + T(P).\end{equation}

First, we explicitly model the computation overhead $W(N, P)$. The $L-N$ layers execute sequentially on the main device, while the $N$ replicated layers are accelerated by parallelism $P$. Thus, the accumulated computation load is:\begin{equation}W(N, P) = (L - N) \cdot t_{comp} + N \cdot \frac{t_{comp}}{P}.\end{equation}

Next, we analyze the communication overhead $T(P)$. A key advantage of \sys{} is that $T(P)$ remains decoupled from the number of layers $N$, as communication occurs exclusively at the boundaries. We model the communication cost based on bandwidth $B$. Let $l$ denote the sequence length, $d_{hidden}$ the hidden dimension, and $bs$ the batch size. The latency is approximated as:\begin{equation}T(P) \approx 2 \cdot \frac{bs/P \cdot (P-1) \cdot l \cdot d_{hidden}}{B}.\end{equation}

Here, the term $bs/P \cdot (P-1)$ denotes the fraction of the batch data offloaded in a scatter-gather operation, and the factor of $2$ accounts for the bidirectional data transfer. Although these phase-dependent variations also influence the per-layer computation time $t_{comp}$, they do not affect the relative speedup ratio calculation within a single token generation step.

Finally, to provide an intuitive metric for performance gains, we define the speedup ratio $S(N, P)$ as the sequential latency divided by the latency under the scaling strategy:\begin{equation}S(N, P) = \frac{L \cdot t_{comp}}{ (L - N)t_{comp} + \frac{N}{P}t_{comp} + T(P) }.\label{eq:speedup_simple}\end{equation}
By introducing the replicated fraction $\alpha(N) = N/L$ and the relative communication cost ratio $\beta(P) = T(P)/(L \cdot t_{comp})$, the formulation can be simplified to reveal the governing dynamics:
\begin{equation}
S(N, P) = \frac{1}{1-\alpha(N)+ \frac{\alpha(N)}{P} + \beta(P)}.
\label{eq:speedup_final}
\end{equation}

This model mathematically validates our design intuition: the speedup is positively correlated with the replicated fraction $\alpha(N)$ and parallelism $P$, yet remains strictly bounded by the fixed communication penalty $\beta(P)$. Crucially, this model confirms that the ``diminishing returns'' observed in Section \ref{sec:ablation} arise as $\alpha(N) \to 1$, where the fixed communication cost $\beta(P)$ starts to dominate the execution time. As an analytical approximation fitted from empirical observations, the model does not aim for cycle-accurate latency prediction. Rather, its value lies in preserving the relative ordering of candidate configurations: the predicted speedup trends are consistent with the distributions observed in the ablation studies. This monotonicity property is sufficient for the scaling algorithm in Section~\ref{sec:algorithm}, where the controller ranks configurations rather than predicting absolute latencies.

\textbf{Generalization to Heterogeneous Environments.}
The simplified model above assumes a homogeneous cluster and perfectly divisible workloads. However, practical deployments often face two sources of heterogeneity: hardware variability (mixed GPU generations~\cite{taco26-lan}) and discrete workload quantization (batch sizes not perfectly divisible by parallelism). To handle these scenarios, we generalize the framework using a configuration vector $\mathbf{v} \in \mathbb{Z}^L$, where the scalar $v^{(l)}$ denotes the parallelism degree for the $l$-th layer. This vector representation captures per-layer heterogeneity and serves as the formal state descriptor for the scaling algorithm presented in Section~\ref{sec:algorithm}.

We define $\mathcal{D}^{(l)}$ as the set of devices assigned to execute layer $l$, with cardinality $|\mathcal{D}^{(l)}| = v^{(l)}$.
When $bs$ is not a multiple of $v^{(l)}$, the load distribution becomes uneven: the device with the heaviest load processes $\lceil bs / v^{(l)} \rceil$ tokens, while others process $\lfloor bs / v^{(l)} \rfloor$. This quantization error introduces a ``straggler effect'' that is amplified when combined with hardware heterogeneity.
Let $t_{comp}^{(k)}$ denote the computation time for device $d_k$ to process a full batch of size $bs$ for a single layer. The generalized computation overhead is the summation of the bottleneck latency across all layers:
\begin{equation}
W(\mathbf{v}) = \sum_{l=1}^{L} \max_{k \in \mathcal{D}^{(l)}} \left( t_{comp}^{(k)} \cdot \frac{\lceil bs / v^{(l)} \rceil}{bs} \right).
\end{equation}

For communication, the transmission rate is bottlenecked by the link with the minimum bandwidth, $B_{min} = \min_{k} B^{(k)}$. Communication occurs at the indices $l$ where the parallelism degree changes (i.e., $v^{(l)} \neq v^{(l+1)}$). The generalized communication cost accounts for the maximum transfer volume required at each transition:
\begin{equation}
T(\mathbf{v}) \approx \sum_{l \in \text{Transitions}} 2 \cdot \frac{ \lceil bs \cdot \frac{v^{(l)}-1}{v^{(l)}} \rceil \cdot d_{hidden} }{B_{min}}.
\end{equation}

The generalized speedup ratio is thus:
\begin{equation}
S(\mathbf{v}) = \frac{\sum_{l=1}^{L} t_{comp}^{(host)}}{ W(\mathbf{v}) + T(\mathbf{v}) }.
\end{equation}
This unified model allows \sys{} to detect subtle performance degradations caused by integer quantization artifacts (e.g., when $bs=1$ and $v^{(l)}>1$, speedup is impossible as $\lceil 1/P \rceil = 1$) and hardware variance, ensuring that the selected strategy $\mathbf{v}$ is physically optimal rather than just theoretically ideal.

\subsection{Initialization Phase}\label{sec:initialization}

To maximize cluster-wide throughput from the outset, \sys{} must first solve a static global placement problem: given a set of model instances $\mathcal{I}$ competing for cluster resources and a pool of underutilized devices, determine the optimal initial layer-wise replication configuration for each instance. We formulate this allocation challenge as a variant of the Multi-Knapsack Problem (MKP), where instances function as utility-generating items and cold devices serve as capacity-constrained knapsacks. Since MKP is NP-hard and exact solutions are impractical for online service bootstrap, \sys{} employs a Priority-based Greedy Heuristic (Algorithm~\ref{alg:initialization}).

\begin{algorithm}[htbp]
   \caption{Initialization: Global Placement}
   \label{alg:initialization}
\begin{algorithmic}[1]
   \STATE {\bfseries Input:} Instances $\mathcal{I}$, Devices $\mathcal{D}$, Current Load $\mathcal{L}_{curr}$, Speedup Model $\mathcal{M}$.
   \STATE {\bfseries Output:} Placement Plan $\mathcal{P}$.
   \STATE $\mathcal{D}_{pool} \gets \{d \in \mathcal{D} \mid \text{Load}(d) < \tau_{idle}\}$
   \FOR{each instance $I_k \in \mathcal{I}$}
       \STATE $\text{Priority}_k \gets \mathcal{L}_{curr, k} - \text{Capacity}(I_k)$
   \ENDFOR
   \STATE $\mathcal{I}_{sorted} \gets \text{Sort}(\mathcal{I}, \text{descending by Priority})$
   \STATE $\mathcal{P} \gets \emptyset$
   \FOR{each $I_k \in \mathcal{I}_{sorted}$}
       \STATE $\mathcal{C}_{feas} \gets \emptyset$
       \FOR{each candidate $(N, P)$ with $1 \le N \le L$, $2 \le P \le P_{max}$}
           \STATE $C_{req} \gets C_{mem}(N, P)$ \quad \textcolor{gray}{// per-device memory from Eq.~\eqref{eq:cost_mem}}
           \STATE $\mathcal{D}_{cand} \gets \{d \in \mathcal{D}_{pool} \mid d.\text{free\_mem} \ge C_{req}\}$
           \IF{$|\mathcal{D}_{cand}| \ge P - 1$}
               \STATE Select any $\mathcal{D}_{sub} \subseteq \mathcal{D}_{cand}$ with $|\mathcal{D}_{sub}| = P - 1$
               \STATE $\mathcal{C}_{feas} \gets \mathcal{C}_{feas} \cup \{(N, P, \mathcal{D}_{sub})\}$
           \ENDIF
       \ENDFOR
       \IF{$\mathcal{C}_{feas} \neq \emptyset$}
           \STATE $(N^*, P^*, \mathcal{D}_{sub}^*) \gets \operatorname{argmax}_{(N,P,\mathcal{D}_{sub}) \in \mathcal{C}_{feas}} S(N, P)$
           \STATE $\mathcal{P} \gets \mathcal{P} \cup \{(I_k, N^*, P^*, \mathcal{D}_{sub}^*)\}$
           \STATE $\mathcal{D}_{pool} \gets \mathcal{D}_{pool} \setminus \mathcal{D}_{sub}^*$
       \ENDIF
   \ENDFOR
   \STATE \textbf{return} $\mathcal{P}$
\end{algorithmic}
\end{algorithm}

The initialization phase proceeds through three sequential steps:

\textbf{Cold Resource Harvesting.} The Monitor component periodically profiles cluster-wide resource utilization via the NVML interface and identifies underutilized devices suitable for repurposing. We define $\text{Load}(d)$ as the GPU compute utilization of device $d$, and introduce a configurable safety threshold $\tau_{idle}$ (empirically set to 70\%). Devices are classified as ``low-load'' candidates and admitted to the dynamic pool $\mathcal{D}_{pool}$ only if $\text{Load}(d) < \tau_{idle}$. This filtering mechanism serves as a primary safety valve, ensuring that resource borrowing remains a non-intrusive supplement rather than disrupting existing workloads.

\textbf{Instance Prioritization.} Given the finite capacity of $\mathcal{D}_{pool}$, prioritizing critical bottlenecks is essential. We define $\text{Capacity}(I_k)$ as the maximum request throughput (RPS) that instance $I_k$ can sustain under its \emph{current} configuration while adhering to the SLO bound. This value is derived from the speedup model in Section~\ref{sec:model}: for a configuration $(N, P)$, the speedup ratio $S(N, P)$ in Equation~\eqref{eq:speedup_final} determines the latency reduction relative to the sequential baseline. By combining this with the SLO constraint and an empirical latency--load relationship (obtained via offline profiling), we compute the maximum sustainable load for each instance. The priority score is: 
\begin{equation}
\text{Priority}_k = \mathcal{L}_{curr, k} - \text{Capacity}(I_k)
\end{equation}
which quantifies the \emph{capacity gap}---the excess load beyond the instance's current serving capability. Sorting instances by descending priority transforms the allocation problem into a sequential decision process, ensuring that instances facing the most acute risk of SLO violation are addressed first.

\textbf{Greedy Allocation and Search Space Tractability.} For each high-priority instance, the algorithm searches for the configuration $(N^*, P^*)$ that maximizes the speedup ratio $S(N, P)$ subject to the physical memory limits of candidate devices. A natural concern is that Line~14 of Algorithm~\ref{alg:initialization} appears to require searching all subsets $\mathcal{D}_{sub} \subseteq \mathcal{D}_{pool}$, which would incur exponential complexity in $|\mathcal{D}_{pool}|$. However, under the homogeneous cluster assumption in Section~\ref{sec:model}, the speedup $S(N, P)$ depends only on the tuple $(N, P)$, not on the specific identity of replica devices. Consequently, the search decomposes into two tractable stages:

\begin{enumerate}
    \item \emph{Configuration enumeration.} The algorithm enumerates all $(N, P)$ pairs where $1 \le N \le L$ and $2 \le P \le P_{max}$. This space contains at most $L \times P_{max}$ candidates---for example, only 256 entries for the Qwen-32B model with $L=64$ and $P_{max}=4$---identical in size to the search space analyzed in Section~\ref{sec:algorithm}.
    \item \emph{Device feasibility check.} For each candidate, the per-device memory requirement is given by $C_{mem}(N, P)$ in Equation~\eqref{eq:cost_mem}. The configuration is feasible if $\mathcal{D}_{pool}$ contains at least $P-1$ devices each satisfying $d.\text{free\_mem} \ge C_{mem}(N, P)$. The device selection is then deterministic: any qualifying subset of size $P-1$ suffices, and \sys{} greedily selects devices with the largest idle memory to minimize fragmentation.
\end{enumerate}

This two-step procedure reduces the complexity from exponential in $|\mathcal{D}_{pool}|$ to $O\big(|\mathcal{I}| \cdot L \cdot P_{max} \cdot |\mathcal{D}_{pool}|\big)$. Evaluating each candidate requires only constant-time arithmetic (computing $C_{mem}$ and $S$) plus a linear scan of $\mathcal{D}_{pool}$, making the per-instance decision latency negligible compared to the service bootstrap interval. 

The algorithm iteratively assigns the best feasible configuration to each instance, removes the allocated devices from $\mathcal{D}_{pool}$, and proceeds to the next instance in the sorted order. Although this greedy strategy does not guarantee a global mathematical optimum, it efficiently approximates the optimal solution and is computationally practical for online serving. In summary, the initialization phase transforms idle cluster resources into scalable replica sets, boosting global system performance while establishing a robust foundation for the dynamic scaling phase described in Section~\ref{sec:algorithm}.

\subsection{Scaling Algorithm}\label{sec:algorithm}

To strictly enforce SLOs while minimizing reconfiguration overhead, we propose a unified layer-wise scaling algorithm (Algorithm \ref{alg:unified_scaling}). Unlike traditional strategies that decouple scale-up and scale-down logic, CoCoScale unifies these operations by treating any significant latency deviation as a trigger for state transition. The Controller continuously monitors the normalized latency deviation $|I.latency - SLO|/SLO$. When this deviation exceeds a tolerance threshold $\delta$, the system initiates a search for the ``nearest'' valid configuration. This approach ensures that whether the system is under-performing (requiring scale-up) or over-provisioned (requiring scale-down), the adjustment target is always to return to the SLO compliance boundary with the minimal necessary operational cost.

\begin{algorithm}[htbp]
   \caption{Unified Layer-wise Scaling}
   \label{alg:unified_scaling}
\begin{algorithmic}[1]
   \STATE {\bfseries Input:} Instances $\mathcal{I}$, Cold Devices $\mathcal{D}_{cold}$, $SLO$, \\ Tolerance $\delta$.
   \WHILE{True}
       \STATE $\mathcal{I}_{scale} \gets \{I \in \mathcal{I} \mid |I.latency - SLO| / SLO > \delta\}$

       \FOR{$I_k \in \mathcal{I}_{scale}$}
           \STATE $S_{curr} \gets S(I_k.N, I_k.P)$
           \STATE $\alpha_{req} \gets S_{curr} \cdot (I_k.latency / SLO)$

           \STATE $\mathcal{C}_{cand} \gets \{(N, P) \mid S(N, P) \ge \alpha_{req} \}$

           \STATE Sort $\mathcal{C}_{cand}$ by $\text{Dist}((I_k.N, I_k.P), (N, P))$

           \STATE $updated \gets \textbf{False}$
           \FOR{$(N^*, P^*) \in \mathcal{C}_{cand}$}
               \STATE $\Delta_{res} \gets \text{Cost}(N^*, P^*) - \text{Cost}(I_k.N, I_k.P)$

               \IF{$\Delta_{res} \le 0$ \OR $\mathcal{D}_{cold}.capacity \ge \Delta_{res}$}
                   \STATE $\textbf{UpdateResource}(\mathcal{D}_{cold}, -\Delta_{res})$
                   \STATE $I_k.\text{config} \gets (N^*, P^*)$
                   \STATE $updated \gets \textbf{True}$; \textbf{break}
               \ENDIF
           \ENDFOR

           \IF{$I_k.latency > SLO$ \textbf{and not} $updated$}
               \STATE $\textbf{TriggerInstanceScale}(I_k)$
           \ENDIF
       \ENDFOR
       \STATE Wait for next interval
   \ENDWHILE
\end{algorithmic}
\end{algorithm}

Using the configuration vector $\mathbf{v}$ defined in Section \ref{sec:model}, we can precisely quantify the ``distance'' between configurations and preserve layer-specific state information. While the tuple $(N, P)$ used in the simplified model summarizes the resource count, it is merely a subset of the information contained in $\mathbf{v}$ (e.g., a configuration $\mathbf{v} = [1,  1, \underbrace{P, \dots, P}_{N}, \dots, 1]^\top$ corresponds to specific layers being replicated). We define the transition cost between the current state $\mathbf{v}_{curr}$ and a target state $\mathbf{v}_{new}$ using the L1 norm distance: $\text{Dist}(\mathbf{v}_{curr}, \mathbf{v}_{new}) = \|\mathbf{v}_{new} - \mathbf{v}_{curr}\|_1$. This metric linearly maps to the data migration overhead, ensuring that the scheduler distinguishes between low-cost adjustments (e.g., extending replication to adjacent layers) and high-cost global reconfigurations.

The scaling process begins by calculating a target speedup ratio $\alpha_{req}$ based on the instance's current speedup and its latency deviation. The system then filters the search space for all candidate vectors $\mathbf{v}_{cand}$ that satisfy the theoretical speedup requirement $S(\mathbf{v}_{cand}) \ge \alpha_{req}$. Crucially, instead of selecting the candidate with the absolute lowest cost, the algorithm sorts candidates by their transition distance $\text{Dist}(\mathbf{v}_{curr}, \mathbf{v}_{cand})$ in ascending order. This optimization strategy prioritizes stability, guiding the system to select the configuration that meets the SLO while introducing the least perturbation to the running inference service.

Finally, the feasibility of the optimal candidate is verified against the available resources in the cluster. We calculate the resource differential $\Delta_{res}$ between the new and current vectors. This unifies the execution logic: if $\Delta_{res} \le 0$ (Scale Down), the transition is automatically approved, and surplus resources are returned to the cold device set $\mathcal{D}_{cold}$; if $\Delta_{res} > 0$ (Scale Up), the transition is executed only if $\mathcal{D}_{cold}$ has sufficient capacity to absorb the increment. As a concrete example, consider an instance currently running with $\mathbf{v}_{curr}$ corresponding to $N=10, P=2$ that experiences a latency spike of 1.3$\times$ the SLO. The controller computes $\alpha_{req}$ and identifies $N=20, P=2$ as the nearest candidate satisfying $S \ge \alpha_{req}$. Since extending replication to 10 adjacent layers requires no additional devices (only memory copies), $\Delta_{res} = 0$ and the transition executes immediately. By strictly adhering to this distance-aware, unified scaling paradigm, CoCoScale effectively eliminates the ``staircase'' provisioning lag, ensuring robust service delivery under volatile workloads.

A natural concern with configuration search is its computational tractability. For a model with $L$ layers and a maximum parallelism degree $P_{max}$, the full configuration space contains $L \times P_{max}$ candidates. In practice, this space remains small: for the Qwen-32B model ($L=64$) with $P_{max}=4$, only 256 entries exist. Since evaluating each candidate requires only constant-time arithmetic (computing $S(N,P)$ and $\text{Dist}$), the per-instance decision latency is negligible compared to the monitoring interval. More importantly, the distance-aware sorting causes the algorithm to terminate early once the first feasible candidate is found, further reducing the effective search in the common case where nearby configurations suffice.

The algorithm converges to a stable configuration when the monitored latency falls within the tolerance band $[SLO \cdot (1-\delta),\; SLO \cdot (1+\delta)]$. Because candidates are sorted by transition distance, consecutive scaling decisions produce monotonically smaller adjustments, which prevents oscillation between distant states.
\section{System Design}\label{sec:sys}

\begin{figure}[htbp]
    \centering
    \includegraphics[width=0.9\linewidth]{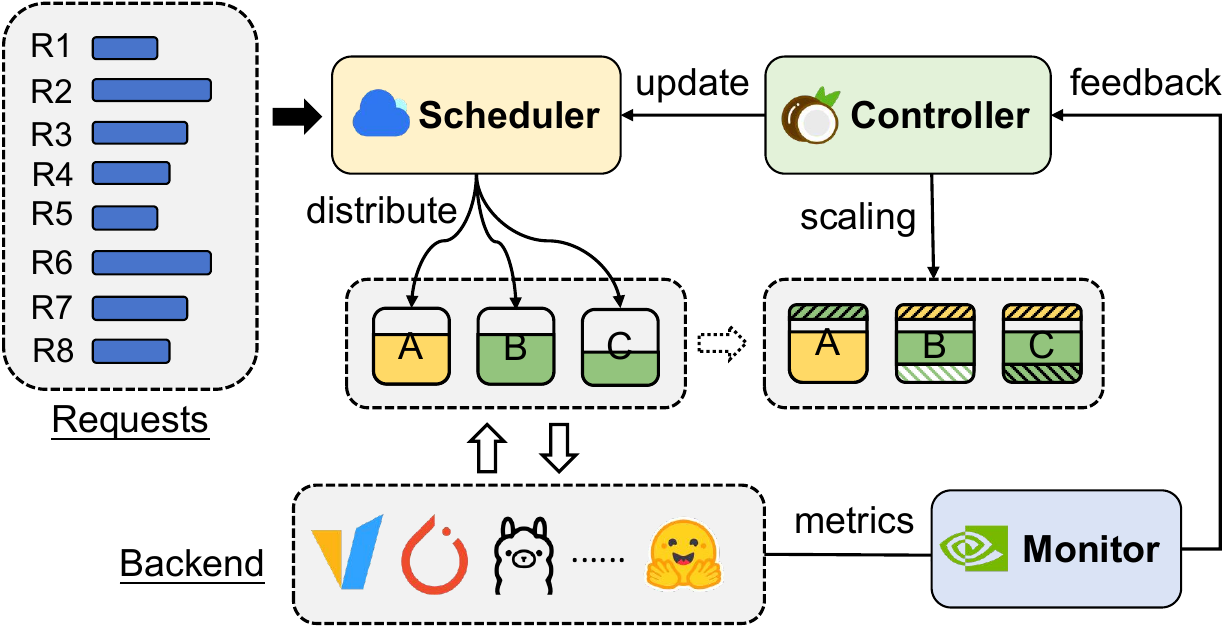}
    \caption{The system overview of \sys{}, illustrating the closed-loop interaction between the Monitor, Controller, and Scheduler for fine-grained resource scavenging.}
    \label{fig:sys}
\end{figure}

To overcome the rigidity of instance-level scaling, we design a hot-pluggable serving framework tailored for \sys{}. As illustrated in Figure \ref{fig:sys}, \sys{} is compatible with mainstream backend engines~\cite{wolf-etal-2020-transformers,xFormers2022,ollama}. It augments the backend via three core components that form a closed-loop control system: the \textit{Monitor} profiles resource availability and detects SLO violations, the \textit{Controller} executes feedback-driven scaling strategies using the analytical model from Section \ref{sec:model}, and the \textit{Scheduler} orchestrates intra-instance request distribution. Together, these components enable continuous monitoring, adaptive decision-making, and seamless resource reallocation.

\subsection{System Architecture}\label{sec:arch}

\textbf{Metrics Monitor.} The \textit{Monitor} extends observation capabilities through two key functions. For resource profiling, it periodically collects hardware metrics (e.g., GPU and memory utilization) via the NVML interface and identifies ``cold'' devices suitable for repurposing based on predefined utilization thresholds. For performance tracking, the Monitor acquires instance-level metrics, including per-request end-to-end latency and throughput, directly from the backend engine. It evaluates SLO compliance by comparing observed statistics against the target latency bound and flags instances whose normalized deviation exceeds the tolerance threshold $\delta$ as candidates for scaling. These aggregated metrics are forwarded to the Controller at each monitoring cycle.

\textbf{Request Scheduler.} The \textit{Scheduler} extends the backend's native dispatching logic to support fine-grained workload distribution~\cite{osdi24-usher}. When an instance operates under a layer-wise DP configuration, the Scheduler partitions incoming request batches into sub-batches according to the active parallelism degree $P$ and dispatches them to the host device and its replicas. By dynamically balancing traffic between the original layers on ``hot'' instances and their replicas on ``cold'' devices, the Scheduler ensures that throughput gains are achieved with minimal synchronization latency, maintaining global cluster load balancing without disrupting the inference critical path.

\textbf{Auto-Scaling Controller.} The \textit{Controller} serves as the central decision engine, leveraging the analytical framework derived in Section \ref{sec:model}. By analyzing aggregated metrics from the Monitor, it determines the optimal $\{N, P\}$ configurations driven by detected SLO violations and resource vacancy. The Controller evaluates candidate configurations using the speedup model $S(N,P)$ and selects the nearest feasible target according to the distance-aware algorithm described in Section \ref{sec:algorithm}. Upon decision finalization, it updates the Scheduler's routing topology and enforces threshold-based safeguards to mitigate performance interference on target devices.

\subsection{Compatibility Discussion}\label{sec:compat}

A practical layer-wise scaling system must be compatible with diverse deployment environments and coexist with the performance optimizations employed by modern inference engines. In this subsection, we discuss four aspects: backend compatibility, parallelism compatibility, multi-tenant interference isolation, and CUDA Graph execution support.

\paragraph{Backend Compatibility.}
\sys{} is designed as a backend-agnostic middleware layer that can be integrated with a variety of LLM serving engines. The layer-wise replication mechanism operates at the model execution level by intercepting the forward pass at configurable layer boundaries, which is a common abstraction across all Transformer-based serving frameworks. Our current prototype is built on Nano-vLLM~\cite{nano-vllm}, but the architecture imposes no engine-specific dependencies. The Monitor communicates with the backend through a standardized metrics interface (latency and throughput counters), and the Scheduler interacts via a request routing API that is orthogonal to the engine's internal scheduling logic. This decoupled design allows \sys{} to be ported to other backends such as vLLM~\cite{sosp23-vllm}, SGLang~\cite{sglang}, or TensorRT-LLM with minimal adaptation, primarily requiring the implementation of a thin adapter layer for metrics collection and batch routing.

\paragraph{Parallelism Compatibility.}
\sys{} adopts layer-wise replication as its core parallelism strategy, and this choice is compatible with the operational constraints of dynamic scaling. In terms of communication efficiency, Tensor Parallelism (TP) requires strictly synchronized AllReduce operations within every layer, inducing heavy overhead that creates bottlenecks on PCIe-based interconnects. In contrast, our strategy restricts data transfer exclusively to the segment boundaries of replicated layers, significantly reducing the frequency of synchronization and making it suitable for scale-out scenarios where bandwidth may be limited. Additionally, implementing dynamic weight resharding for TP at runtime incurs prohibitive engineering complexity. Pipeline Parallelism (PP) is primarily designed to improve throughput via micro-batching but inevitably introduces ``bubble'' latency, which counteracts our goal of minimizing real-time inference delay. Our layer-replication method offers a favorable trade-off by simplifying the control plane to a manageable memory copy operation while delivering predictable speedups. Furthermore, layer-wise replication is compositionally compatible with existing static parallelism configurations: an instance already using intra-node TP can additionally employ layer-wise replication across nodes for dynamic capacity expansion, as the two mechanisms operate on orthogonal axes.

\paragraph{Interference Isolation.}
To evaluate the isolation capabilities of \sys{} in multi-tenant environments, we conducted stress tests focusing on the interference caused by scaling operations on co-located workloads. We measured the performance impact on adjacent active instances while simultaneously triggering scaling events (weight migration and KV-cache transfer) on neighboring devices within the same node. Experimental data indicates that during dynamic migration, the throughput fluctuation of adjacent active instances is suppressed to less than 3\%. We also verified that the outputs produced by \sys{} instances remain bit-wise consistent with the baseline, ensuring strict correctness guarantees.

A critical observation concerns the management of ``Cold Instances,'' which are replicas that are provisioned but currently idle. While these instances occupy GPU memory (making immediate release difficult without process termination), their impact on the performance of active tasks is negligible. Since cold instances do not compete for computation cycles (SMs), the scaling process primarily utilizes P2P bandwidth. Our ring-based flow control effectively limits this bandwidth usage, ensuring that state migration does not disturb the inference stability of neighboring high-priority services.

\paragraph{CUDA Graph Execution Compatibility.}\label{sec:graph}
To strictly enforce low-latency objectives during the decoding phase, \sys{} is designed to be fully compatible with CUDA Graphs. While dynamic scaling typically conflicts with the static nature of computation graphs (which require fixed tensor shapes and execution paths), our layer-wise replication strategy circumvents this limitation by treating each replica as an independent, static execution unit.

In contrast to TP, which requires complex communication group updates during scaling, our replicas execute a fixed set of continuous layers. This allows us to capture the forward pass of each replica into a CUDA Graph at initialization. During inference, the runtime controller simply routes different sub-batches to these pre-captured graphs. To handle the variable sequence lengths in the KV-cache without breaking the graph, we employ a pre-allocation strategy with pointer arithmetic, ensuring that the tensor memory addresses remain constant across iterations.

Graph re-capture is only triggered when the scaling algorithm alters the configuration vector $\mathbf{v}$ (i.e., changing the number of replicated layers $N$ or parallelism $P$). To minimize the overhead of these transition events, we implement a persistent memory pool mechanism. When a new graph configuration is required, the system reuses the existing allocated memory buffers instead of triggering expensive device memory reallocation. The resulting re-capture overhead is evaluated in Section~\ref{sec:overhead}. Consequently, \sys{} enjoys the kernel-launch-free performance of static graphs while maintaining the elasticity of dynamic scheduling.
\section{Perforamnce Evaluations}
\label{sec:evaluation}


\subsection{Experimental Setup}

\textbf{Testbed.} Our experiments are conducted on a cluster with 4 NVIDIA H20 GPUs. \sys{} is built on Nano-vLLM~\cite{nano-vllm} and utilizes NVLink for rapid layer-wise weight and KV cache migration~\cite{qin2024mooncakekvcachecentricdisaggregatedarchitecture}.

\textbf{Workload Traces.} We evaluate \sys{} on two production request traces that differ in intensity and burstiness: (1) \textbf{Alibaba Trace:} A 60-minute production trace from Alibaba's online serving platform with moderate burstiness (average RPS: \textbf{1.62}, peak RPS: \textbf{11}). (2) \textbf{Azure Trace \cite{azure2024public}:} A longer 130-minute trace with higher load intensity and more severe burstiness (average RPS: \textbf{7.09}, peak RPS: \textbf{23}), representing a more challenging production scenario. Both traces exhibit pronounced temporal variability, consistent with prior characterizations~\cite{arxiv24-burstgp}. To preserve privacy while retaining long-context behavior, we use prompts from LongBench~\cite{bai2024longbench2}.

\textbf{Multi-Instance Configuration.} We deploy 4 concurrent model instances competing for 4 GPUs under skewed workload distributions. Processing identical traces simultaneously creates contention where underutilized (cold) instances relinquish resources to overloaded (hot) instances.

\textbf{Baselines.} We compare \sys{} against two state-of-the-art baselines:
(1) \textbf{vLLM (Static)}~\cite{sosp23-vllm}, which provisions resources based on average load and assigns each engine a fixed 1-GPU allocation; and
(2) \textbf{Alibaba Autoscaler}, a utilization-threshold policy that samples resource usage every \textbf{1\,s}, scales out when utilization exceeds \textbf{80\%}, and scales in when utilization drops below \textbf{30\%} (no cooling period), and these values are widely used as scaling thresholds in realistic production systems. 
This baseline employs a \emph{warm pool} strategy where scaled-in instances are retained rather than terminated, mitigating cold-start overhead (up to 241s, as shown in Figure~\ref{fig:combined_overhead}b).

\begin{table}[htbp]
\centering
\caption{Performance comparison across Alibaba (60min, average RPS is 1.62, SLO=30s) and Azure (130min, average RPS is 7.09, SLO=27s) traces.}
\label{tab:main_results}
\resizebox{0.8\columnwidth}{!}{%
\small
\begin{tabular}{clcccccc}
\toprule
& & \multicolumn{3}{c}{\textbf{Alibaba Trace}} & \multicolumn{3}{c}{\textbf{Azure Trace}} \\
\cmidrule(lr){3-5} \cmidrule(lr){6-8}
\textbf{Model} & \textbf{Method} & \textbf{Avg (s)} & \textbf{P99 (s)} & \textbf{SLO Met} & \textbf{Avg (s)} & \textbf{P99 (s)} & \textbf{SLO Met} \\
\midrule
\multirow{3}{*}{Qwen3-8B} 
& vLLM (Static) & $39.7$ & $50.1$ & $0.7\%$ & $34.2$ & $40.7$ & $1.1\%$ \\
& Alibaba & $30.7$ & $35.3$ & $40.6\%$ & $32.2$ & $37.2$ & $0.8\%$ \\
\rowcolor{blue!10}
& \textbf{\sys{}} & \textbf{22.1} & \textbf{28.2} & \textbf{100\%} & \textbf{17.5} & \textbf{23.5} & \textbf{100\%} \\
\midrule
\multirow{3}{*}{Qwen3-14B} 
& vLLM (Static) & $39.5$ & $49.6$ & $0.8\%$ & $33.5$ & $41.1$ & $1.3\%$ \\
& Alibaba & $30.3$ & $35.2$ & $48.3\%$ & $31.9$ & $37.1$ & $2.4\%$ \\
\rowcolor{blue!10}
& \textbf{\sys{}} & \textbf{21.9} & \textbf{27.9} & \textbf{100\%} & \textbf{18.1} & \textbf{24.1} & \textbf{100\%} \\
\midrule
\multirow{3}{*}{Qwen3-32B} 
& vLLM (Static) & $39.3$ & $48.3$ & $1.0\%$ & $29.4$ & $36.0$ & $21.1\%$ \\
& Alibaba & $27.5$ & $32.1$ & $90.7\%$ & $31.2$ & $36.4$ & $2.4\%$ \\
\rowcolor{blue!10}
& \textbf{\sys{}} & \textbf{21.8} & \textbf{27.6} & \textbf{100\%} & \textbf{17.0} & \textbf{23.6} & \textbf{100\%} \\
\bottomrule
\end{tabular}
}
\end{table}

\subsection{Main Results and Analysis}
\begin{figure*}[htbp]
    \centering
    \begin{minipage}{0.32\textwidth}
        \centering
        \includegraphics[width=\linewidth]{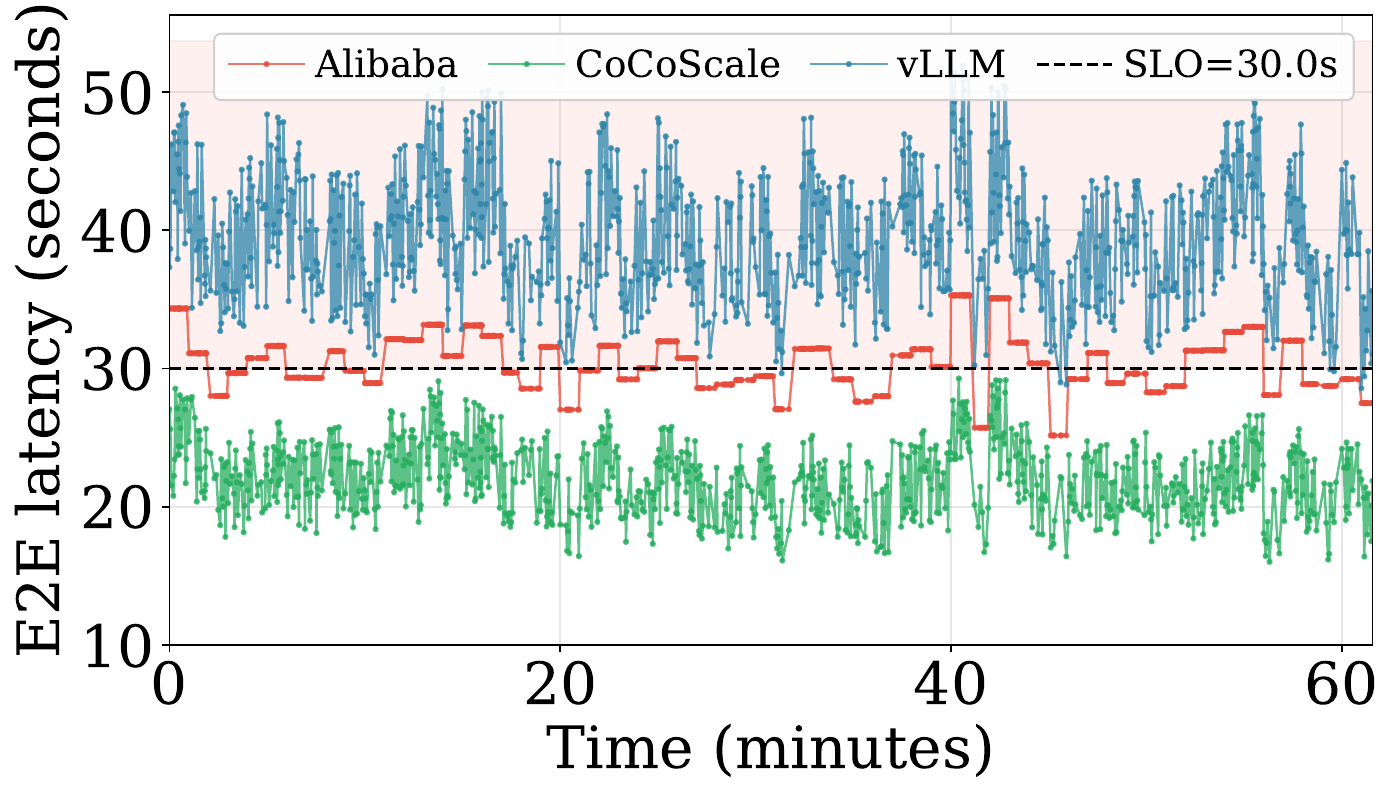}
        \small (a) Qwen3-8B (Alibaba)
    \end{minipage}
    \begin{minipage}{0.32\textwidth}
        \centering
        \includegraphics[width=\linewidth]{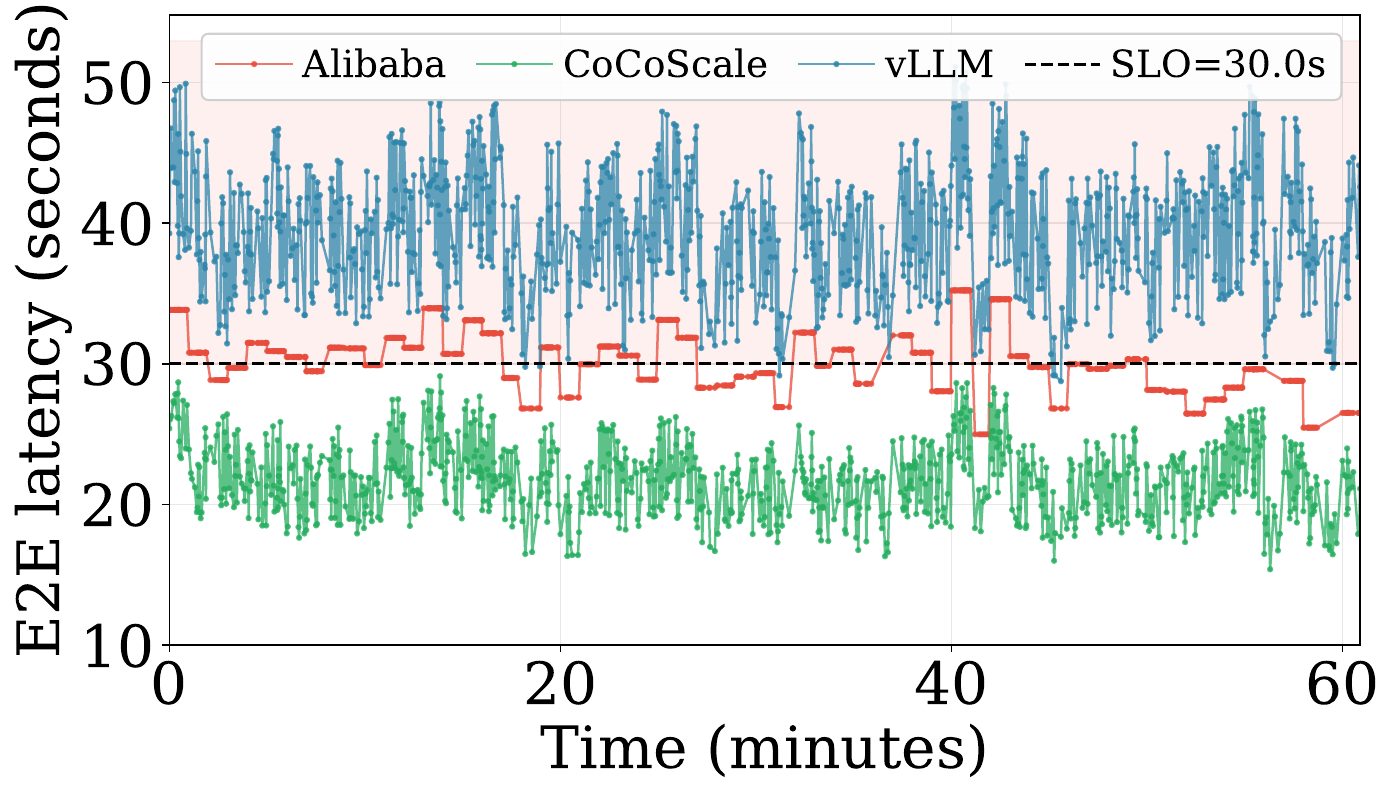}
        \small (b) Qwen3-14B (Alibaba)
    \end{minipage}
    \begin{minipage}{0.32\textwidth}
        \centering
        \includegraphics[width=\linewidth]{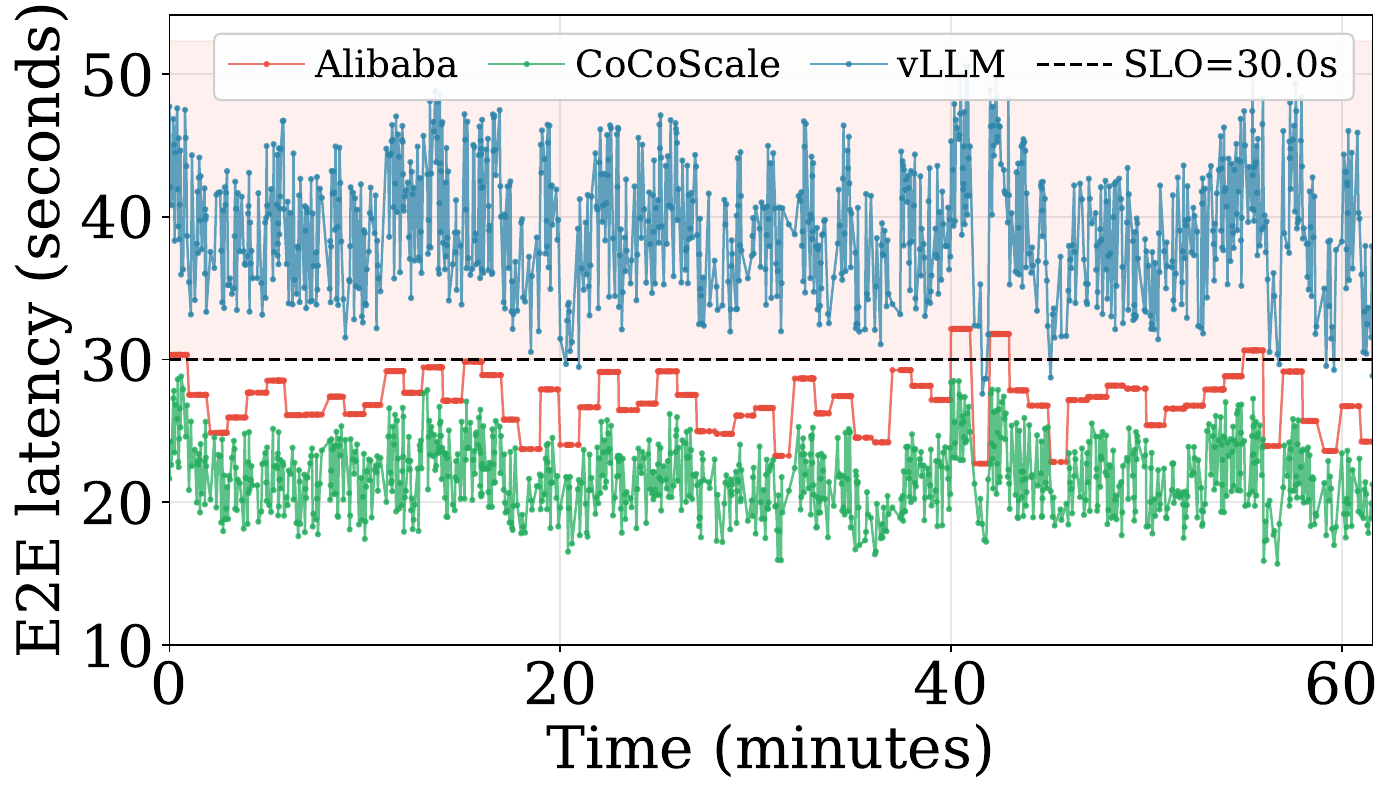}
        \small (c) Qwen3-32B (Alibaba)
    \end{minipage} \\
    \begin{minipage}{0.32\textwidth}
        \centering
        \includegraphics[width=\linewidth]{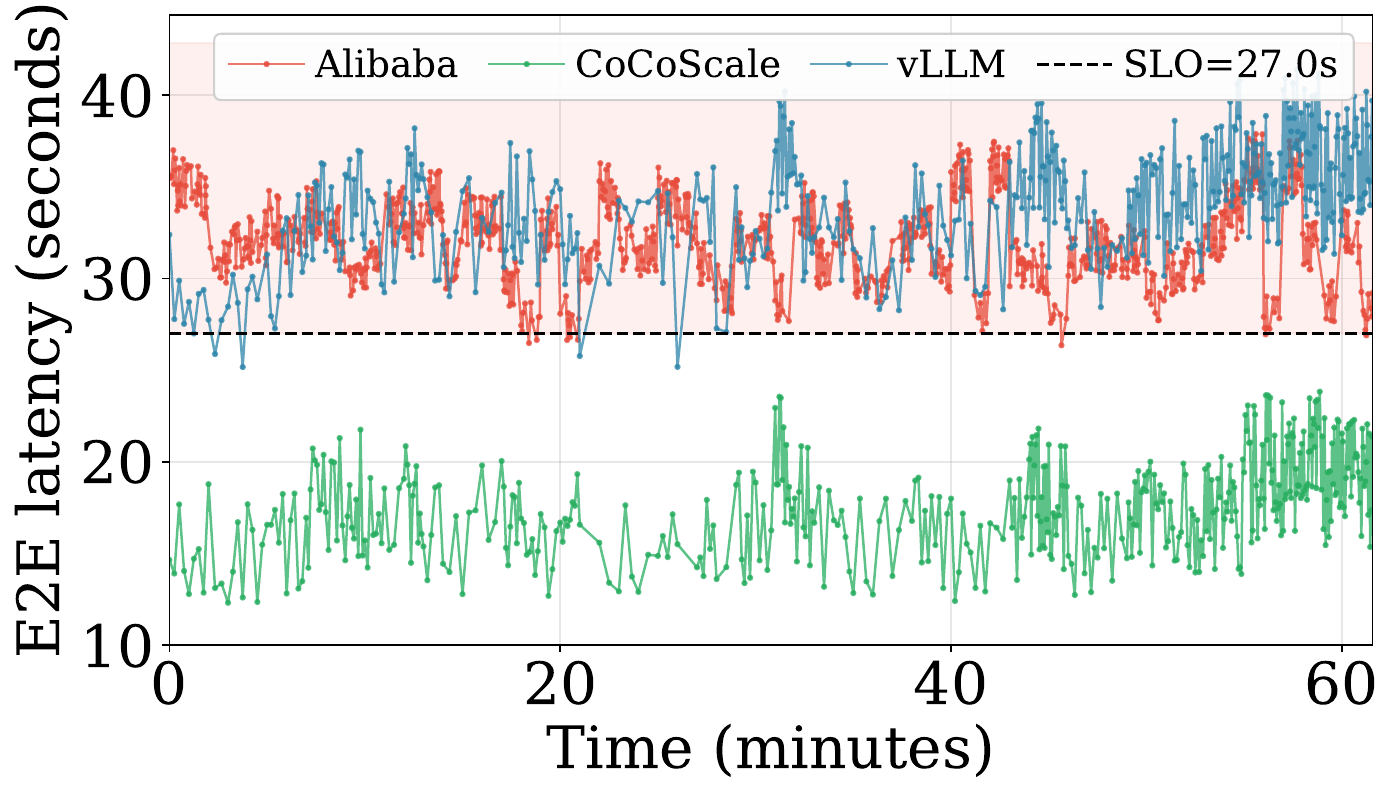}
        \small (d) Qwen3-8B (Azure)
    \end{minipage}
    \begin{minipage}{0.32\textwidth}
        \centering
        \includegraphics[width=\linewidth]{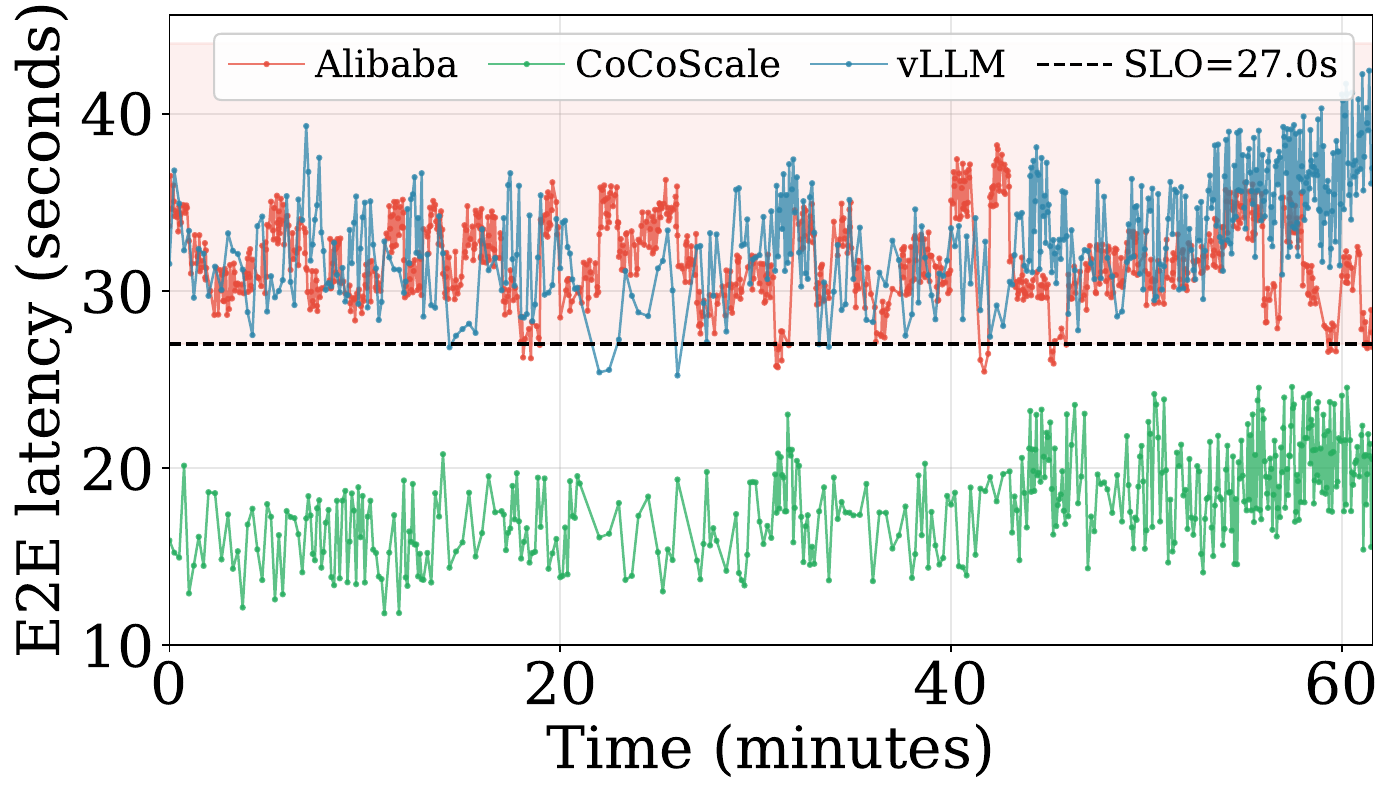}
        \small (e) Qwen3-14B (Azure)
    \end{minipage}
    \begin{minipage}{0.32\textwidth}
        \centering
        \includegraphics[width=\linewidth]{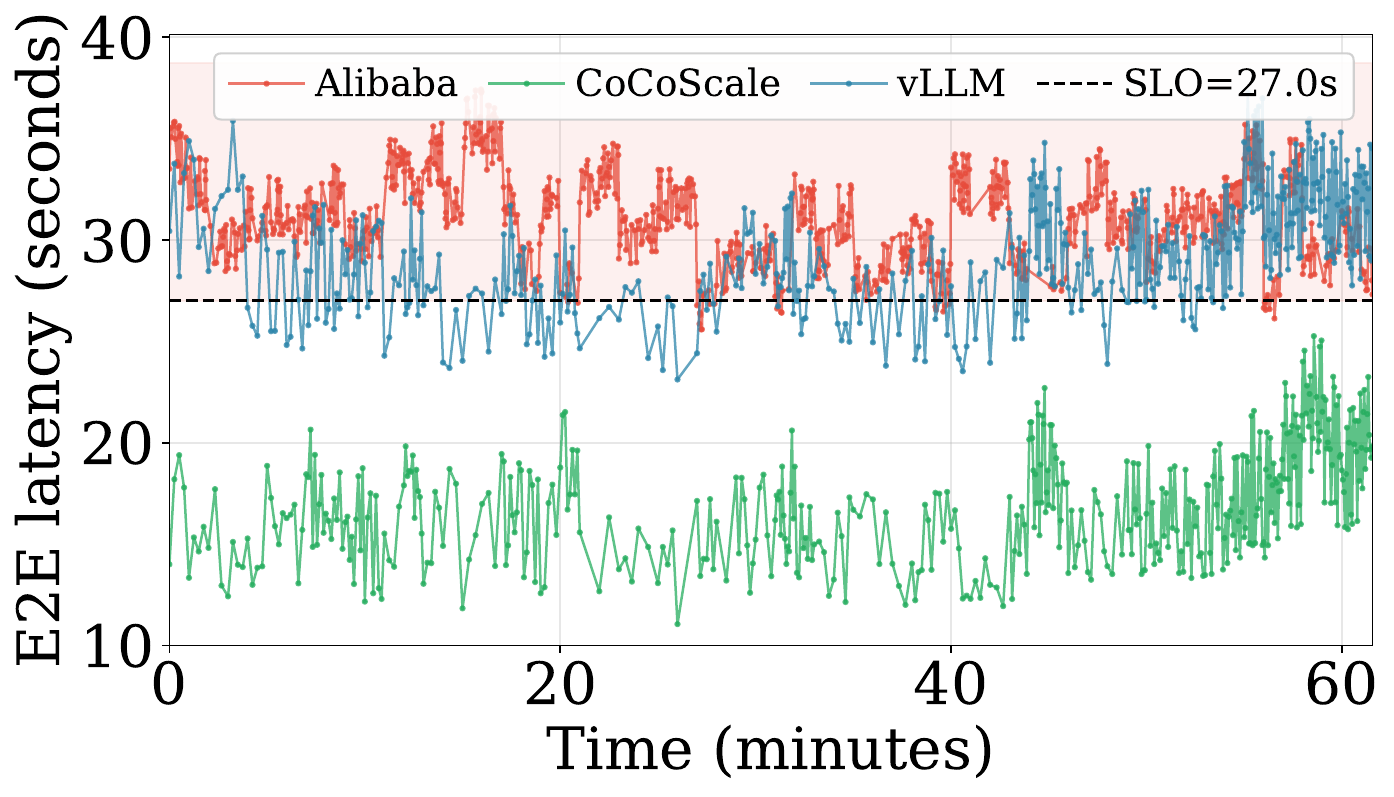}
        \small (f) Qwen3-32B (Azure)
    \end{minipage}
    \caption{Temporal latency comparison across production traces. \textbf{Top (a--c):} Alibaba trace (60 min, moderate load). \textbf{Bottom (d--f):} Azure trace (130 min, 4.4$\times$ higher load). \sys{} maintains stable latency through adaptive layer-wise scaling, while baselines exhibit latency spikes during demand surges due to resource contention (vLLM Static) and reconfiguration delays (Alibaba Autoscaler).}
    \label{fig:temporal_results}
\end{figure*}

\textbf{Latency Performance.}
As shown in Table~\ref{tab:main_results} and Figure~\ref{fig:temporal_results}, \sys{} consistently outperforms both baselines while maintaining temporal stability across both traces.
Under the moderate Alibaba trace (SLO=30\,s), CoCoScale maintains low latency (green lines in Figures~\ref{fig:temporal_results}a--c) throughout the 60-minute window, whereas vLLM (Static) (blue) and Alibaba autoscaler (red) exhibit significant spikes exceeding the SLO threshold during demand surges.
Similarly, under the intensive Azure trace (4.4$\times$ higher load, SLO=27\,s), CoCoScale demonstrates robust stability (Figures~\ref{fig:temporal_results}d--f), while baselines suffer prolonged high latency due to resource contention (vLLM Static) and reconfiguration delays (Alibaba autoscaler).
Quantitatively, \sys{} achieves \textbf{20.7\%--28.1\%} lower average latency than Alibaba autoscaler (P99 latency \textless \, \textbf{28.3\,s}) on the Alibaba trace, and \textbf{43.3\%--45.7\%} reduction on the Azure trace.

\textbf{Per-Model Analysis.}
Examining the results across model sizes reveals that \sys{} provides consistent improvements regardless of model complexity. For Qwen3-8B, the smallest model, \sys{} reduces the average latency from 30.7\,s (Alibaba autoscaler) to 22.1\,s on the Alibaba trace, a 28.0\% improvement. The relatively lower computational demand of Qwen3-8B means that layer-wise replication can cover a larger fraction of the model's layers within the available memory budget, yielding a higher effective speedup ratio. For Qwen3-14B, we observe similar gains (27.7\% reduction), with the P99 latency dropping to 27.9\,s. The Qwen3-32B model is the most demanding in terms of both computation and memory. Despite this, \sys{} achieves a 20.7\% reduction in average latency on the Alibaba trace and an even more pronounced 45.5\% reduction on the Azure trace. This is because under high concurrency, the larger batch sizes make the parallelism gains from layer-wise DP more effective, as the communication overhead ($\beta(P)$ in our model) is amortized over a larger number of tokens.

\textbf{Trace Comparison.}
The performance gap between \sys{} and the baselines widens significantly under the Azure trace. This trace exhibits a 4.4$\times$ higher average RPS and more severe burstiness, with peak loads reaching 23 RPS compared to 11 RPS in the Alibaba trace. Under such conditions, the Alibaba autoscaler's SLO attainment drops from 40.6\%--90.7\% to a mere 0.8\%--2.4\%, because its utilization-threshold scaling policy cannot react quickly enough to absorb the sudden load spikes. The cold-start latency of provisioning new instances (74--241\,s as shown in Figure~\ref{fig:combined_overhead}b) far exceeds the duration of individual burst events. In contrast, \sys{} completes layer-wise scaling in 1--2 seconds, allowing it to preemptively expand capacity before the burst saturates the existing resources. This advantage is most visible in Figures~\ref{fig:temporal_results}d--f, where the green line (\sys{}) remains consistently below the SLO threshold even during the most intense burst periods around the 40--60 minute marks.

\textbf{SLO Attainment.}
These latency improvements translate directly into \textbf{100\% SLO attainment} for \sys{} across all models and both traces (Table~\ref{tab:main_results}). In stark contrast, the Alibaba autoscaler achieves only \textbf{40.6\%--90.7\%} on the moderate Alibaba trace, and catastrophically drops to merely \textbf{0.8\%--2.4\%} under the high-intensity Azure trace. vLLM (Static) performs worst with only \textbf{0.7\%--21.1\%} attainment, rendering it unsuitable for production scenarios with strict latency requirements. The full SLO compliance of \sys{} is a direct consequence of the fine-grained elasticity: by adjusting capacity in fractional increments (e.g., replicating 10 additional layers rather than a full instance), the system can precisely match the target speedup ratio without over-provisioning, maintaining tight adherence to the SLO boundary throughout the entire serving period.



\subsection{Overhead Analysis}
\label{sec:overhead}

To validate system agility, we evaluate the setup overhead of \sys{}'s layer-wise scaling compared to traditional full-instance scaling.

\begin{figure}[htbp]
    \centering
    \begin{subfigure}[t]{0.45\linewidth}
        \centering
        \includegraphics[width=\linewidth]{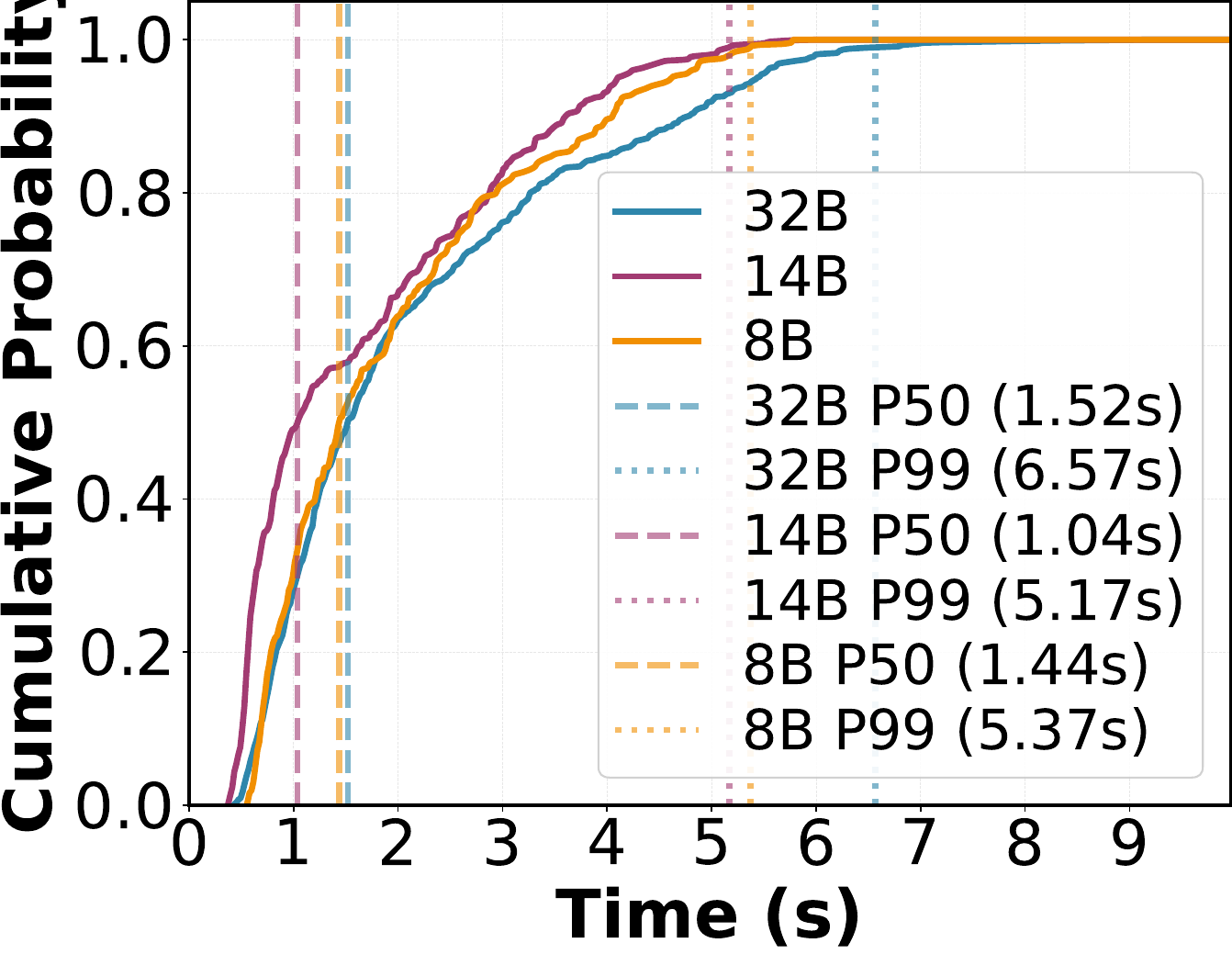}
        \caption{CDF of \sys{} scale-up latency. P50 stays stable (1.04\,s to 1.52\,s) across model sizes with consistently low tail latency.}
        \label{fig:cdf_overhead}
    \end{subfigure}
    \hfill
    \begin{subfigure}[t]{0.48\linewidth}
        \centering
        \includegraphics[width=\linewidth]{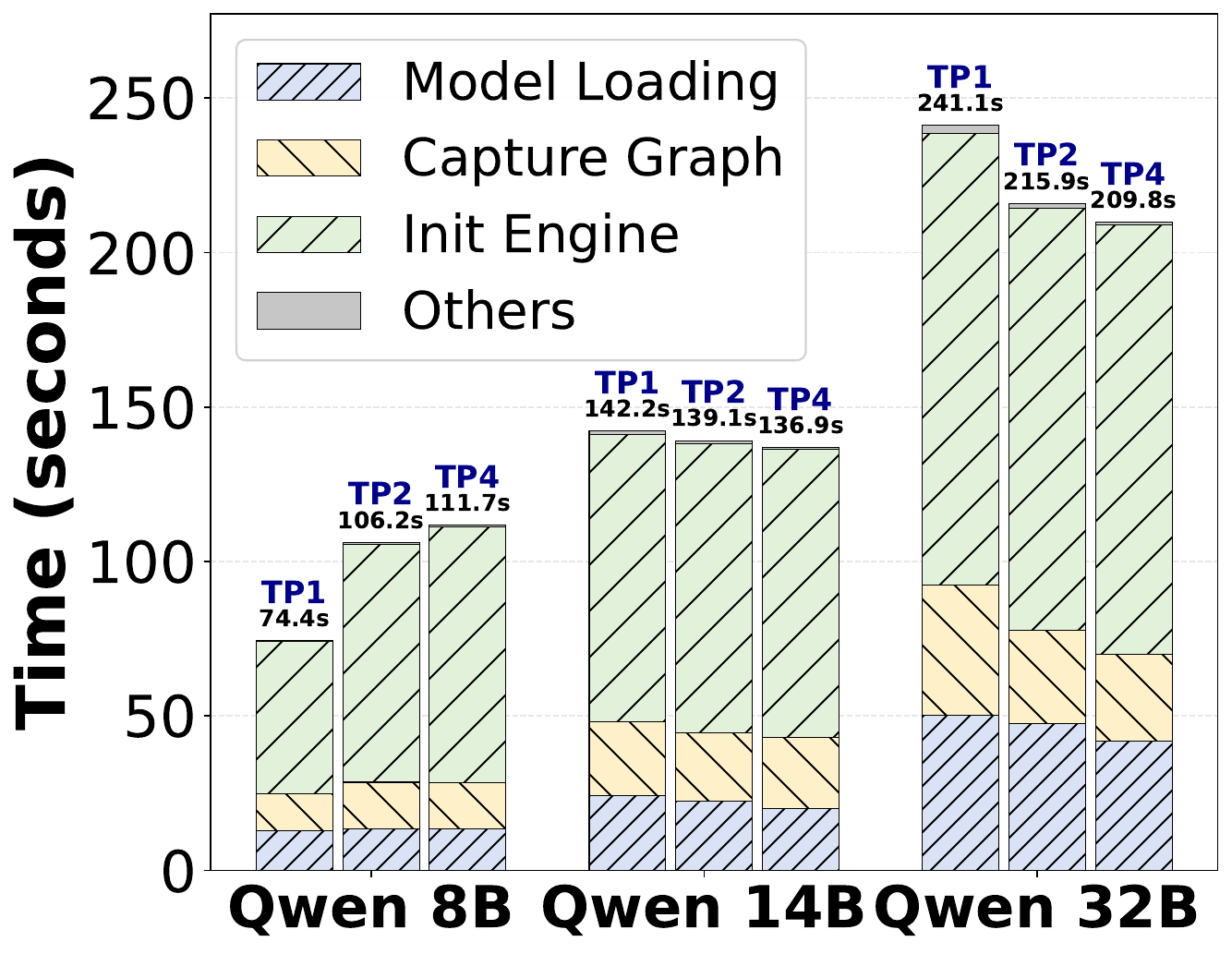}
        \caption{Startup breakdown of vLLM. End-to-end delay (74\,s to 241\,s) is dominated by model loading and engine initialization.}
        \label{fig:vllm_bar}
    \end{subfigure}
    \caption{Scaling overhead comparison. (a) \sys{} agility. (b) Traditional scaling bottlenecks.}
    \label{fig:combined_overhead}
\end{figure}

\textbf{\sys{} Scaling Agility.} 
As shown in Figure~\ref{fig:cdf_overhead}, \sys{} achieves sub-second to second-level response times. For the Qwen3-32B model, the \textbf{P50 scaling latency is 1.52s}, which is nearly invariant to model size. This is enabled by our ring-multicast protocol, which migrates only essential layers and KV caches, avoiding full engine re-initialization.

\textbf{Bottlenecks in Traditional Scaling.} 
In contrast, Figure~\ref{fig:vllm_bar} highlights the prohibitive costs of full-instance scaling. A Qwen3-32B instance requires up to \textbf{241.1s} to become operational, with over 80\% of the time spent on model loading and CUDA graph capturing. Compared to this baseline, \sys{} reduces the scaling latency by 97.9\%--99.3\% , enabling the system to react effectively to rapid production workload fluctuations.

\textbf{Sensitivity to Interconnect.} While our current evaluation utilizes NVLink for high-speed layer migration (PCIe 5.0 equivalent bandwidth), we acknowledge that \sys{}'s performance may vary under slower interconnects (e.g., PCIe 4.0 or Ethernet). The layer-wise migration latency would increase proportionally with reduced bandwidth, potentially affecting the agility demonstrated in Figure~\ref{fig:cdf_overhead}. We leave the comprehensive sensitivity analysis of migration overhead under diverse network topologies to future work.


\section{Related Work}
\label{sec:related}

\paragraph{Instance-level Scaling and Cold Start.} Conventional scaling strategies incur significant cold-start latencies due to full-model loading. ServerlessLLM~\cite{ServerlessLLM} accelerates checkpoint loading through locality-aware scheduling, and BlitzScale~\cite{BlitzScale} reduces startup time by pre-warming containers and overlapping initialization stages. HydraServe~\cite{lou2025hydraserve} further improves cold-start performance by sharing model weights across co-located instances. Yu et al.~\cite{taco26-yu} address the serverless inference cold-start problem through GPU-efficient model swapping, enabling low-latency instance transitions. Despite these optimizations, all approaches remain restricted to coarse-grained instance boundaries, where the fundamental unit of scaling is a complete model replica. This granularity mismatch prevents them from responding to sub-second traffic bursts without significant over-provisioning. \sys{} diverges from this paradigm by introducing layer-wise scaling, which decouples capacity expansion from model loading to achieve near-instantaneous responsiveness.

\paragraph{Resource Pooling and Multiplexing.} Systems like Aegaeon~\cite{Aegaeon} and MuxServe~\cite{duan2024muxserve} improve utilization through model swapping or live migration but incur prohibitive reconfiguration overheads during sudden load shifts. These approaches typically require moving heavy model states, which blocks inference during critical bursts. Llumnix~\cite{osdi24-llumnix} introduces live migration of inference contexts across GPUs to improve load balancing but still operates at the instance level with significant migration overhead for large models. In contrast, \sys{} treats cold resources as a hot-pluggable pool, allowing hot instances to seamlessly borrow compute cycles without displacing resident models or triggering heavy state migration.

\paragraph{Parallelism Granularity and Serving.} Existing frameworks often rely on static parallelism configurations~\cite{osdi23-alpaserve,LoongServe,choi2022serving} or phase disaggregation~\cite{osdi24-distserve,Patel2023SplitwiseEG} that require complex global synchronization to adjust at runtime. Similarly, changing Tensor~\cite{arxiv19-megatron} or Pipeline Parallelism~\cite{wan2025pipeoffload} degrees necessitates expensive reshuffling of weights and KV caches. Recent work such as FlexPipe~\cite{lin2025flexpipe} explores adaptive pipeline configurations but remains constrained by the pipeline bubble problem. Wei et al.~\cite{taco26-wei} propose interleaved parallelism to dynamically balance latency and throughput in distributed inference, but their approach operates at the parallelism configuration level rather than the layer level. \sys{} introduces layer-wise DP to enable flexible, continuous capacity adjustments that precisely align with dynamic demand curves without rigid architectural constraints.

\paragraph{Disaggregated Inference and KV Cache Management.} A growing body of work advocates disaggregating the prefill and decode phases of LLM inference onto separate hardware to improve resource efficiency. Splitwise~\cite{Patel2023SplitwiseEG} and DistServe~\cite{osdi24-distserve} partition inference requests such that compute-intensive prefill runs on high-throughput accelerators while memory-bound decode runs on capacity-optimized devices. Complementing this direction, efficient KV cache management has emerged as a critical concern. Mooncake~\cite{qin2024mooncakekvcachecentricdisaggregatedarchitecture} proposes a KV-cache-centric disaggregated architecture that decouples cache storage from computation, while vLLM~\cite{sosp23-vllm} introduces paged attention to reduce memory fragmentation. While \sys{} currently operates within a non-disaggregated setting, its layer-wise replication mechanism is orthogonal to these techniques: the fine-grained cache partitioning used in our scatter-gather protocol shares design principles with paged attention, and extending layer-wise scaling to disaggregated clusters is a promising direction for future work.

\section{Conclusion}
We propose \sys{} to overcome the latency and inefficiency of coarse-grained scaling by introducing fine-grained elasticity through layer-wise data parallelism. The core insight is that the layered structure of Transformer models provides a natural decomposition point for scaling: by selectively replicating subsets of layers onto idle resources, \sys{} achieves non-integer capacity expansion without full-model redeployment. We developed an analytical model that captures the interplay between the replicated layer count $N$, parallelism degree $P$, and communication overhead, enabling the system to determine cost-effective configurations in real time. Built on this foundation, a unified scaling algorithm dynamically adjusts configurations to maintain SLO compliance with minimal reconfiguration cost.

Evaluations on production traces from Alibaba and Azure demonstrate that \sys{} reduces scale-up latency by 97.9\%--99.3\% compared to traditional instance-level scaling. Under both moderate and high-intensity workloads, \sys{} achieves 100\% SLO attainment while reducing average end-to-end latency by 20.7\% to 28.1\%. These results confirm that decoupling capacity adjustment from rigid instance boundaries enables precise resource alignment with dynamic demand.

\bibliographystyle{ACM-Reference-Format}
\bibliography{ref}

@article{taco26-wei,
author = {Wei, Jinhui and Cheng, Shenggan and Zhu, Wei and Jiang, Jiazhi and Huang, Dan and Chen, Zhiguang and Du, Jiangsu and Lu, Yutong},
title = {Dynamic Latency-Throughput Balancing in Distributed Large Model Inference with Interleaved Parallelism},
year = {2026},
volume = {23},
number = {1},
issn = {1544-3566},
doi = {10.1145/3797040},
journal = {ACM Trans. Archit. Code Optim.},
month = mar,
articleno = {27},
numpages = {26}
}

@misc{xu2026cloudnativedistributedsystemsefficient,
      title={Cloud-native and Distributed Systems for Efficient and Scalable Large Language Models -- A Research Agenda}, 
      author={Minxian Xu and Jingfeng Wu and Shengye Song and Satish Narayana Srirama and Bahman Javad and Rajiv Ranjan and Devki Nandan Jha and Sa Wang and Wenhong Tian and Huanle Xu and Li Li and Zizhao Mo and Shuo Ren and Thomas Kunz and Petar Kochovski and Vlado Stankovski and Kejiang Ye and Chengzhong Xu and Rajkumar Buyya},
      year={2026},
      eprint={2604.17227},
      archivePrefix={arXiv},
      primaryClass={cs.DC},
      url={https://arxiv.org/abs/2604.17227}, 
}

@article{taco26-lan,
author = {Lan, Hao and Zhou, Ziang and Zhu, Qi and Yan, Wei and Hao, Qinfen and Ye, Xiaochun and Liu, Yong and Sun, Ninghui},
title = {Heterogeneous Confidential Computing System for Large Language Models: A Survey},
year = {2026},
volume = {23},
number = {1},
issn = {1544-3566},
doi = {10.1145/3779307},
journal = {ACM Trans. Archit. Code Optim.},
month = mar,
articleno = {4},
numpages = {26}
}

@article{taco26-yu,
author = {Yu, Minchen and Wang, Ao and Wu, Bohui and Liu, Yuxuan and Chen, Dong and Yu, Haoxuan and Wang, Wei and Chen, Ruichuan and Nie, Dapeng and Yang, Haoran and Ding, Yu},
title = {Enabling Low-Latency, GPU-Efficient Serverless Inference with Model Swapping},
year = {2026},
issn = {1544-3566},
doi = {10.1145/3800690},
journal = {ACM Trans. Archit. Code Optim.},
month = apr
}

@inproceedings{sglang,
  title={Sglang: Efficient execution of structured language model programs},
  author={Zheng, Lianmin and Yin, Liangsheng and Xie, Zhiqiang and Sun, Chuyue and Huang, Jeff and Yu, Cody H and Cao, Shiyi and Kozyrakis, Christos and Stoica, Ion and Gonzalez, Joseph E and others},
  journal={Advances in neural information processing systems},
  volume={37},
  pages={62557--62583},
  year={2024}
}

@inproceedings{osdi24-llumnix,
author = {Biao Sun and Ziming Huang and Hanyu Zhao and Wencong Xiao and Xinyi Zhang and Yong Li and Wei Lin},
title = {Llumnix: Dynamic Scheduling for Large Language Model Serving},
booktitle = {18th USENIX Symposium on Operating Systems Design and Implementation (OSDI 24)},
year = {2024},
isbn = {978-1-939133-40-3},
address = {Santa Clara, CA},
pages = {173--191},

publisher = {USENIX Association},
month = jul
}

@article{bai2024longbench2,
  title={LongBench v2: Towards Deeper Understanding and Reasoning on Realistic Long-context Multitasks}, 
  author={Yushi Bai and Shangqing Tu and Jiajie Zhang and Hao Peng and Xiaozhi Wang and Xin Lv and Shulin Cao and Jiazheng Xu and Lei Hou and Yuxiao Dong and Jie Tang and Juanzi Li},
  journal={arXiv preprint arXiv:2412.15204},
  year={2024}
}

@inproceedings{osdi24-distserve,
author = {Yinmin Zhong and Shengyu Liu and Junda Chen and Jianbo Hu and Yibo Zhu and Xuanzhe Liu et al.},
title = {DistServe: Disaggregating Prefill and Decoding for Goodput-optimized Large Language Model Serving},
booktitle = {18th USENIX Symposium on Operating Systems Design and Implementation (OSDI 24)},
year = {2024},
isbn = {978-1-939133-40-3},
address = {Santa Clara, CA},
pages = {193--210},

publisher = {USENIX Association},
month = jul
}

@inproceedings{osdi24-usher,
author = {Sudipta Saha Shubha and Haiying Shen and Anand Iyer},
title = {USHER: Holistic Interference Avoidance for Resource Optimized {ML} Inference},
booktitle = {18th USENIX Symposium on Operating Systems Design and Implementation (OSDI 24)},
year = {2024},
isbn = {978-1-939133-40-3},
address = {Santa Clara, CA},
pages = {947--964},

publisher = {USENIX Association},
month = jul
}

@inproceedings{sosp23-vllm,
author = {Kwon, Woosuk and Li, Zhuohan and Zhuang, Siyuan and Sheng, Ying and Zheng, Lianmin et al.},
title = {Efficient Memory Management for Large Language Model Serving with PagedAttention},
year = {2023},
isbn = {9798400702297},
publisher = {Association for Computing Machinery},
address = {New York, NY, USA},
doi = {10.1145/3600006.3613165},
booktitle = {Proceedings of the 29th Symposium on Operating Systems Principles},
pages = {611–626},
numpages = {16},
location = {Koblenz, Germany},
series = {SOSP '23}
}

@inproceedings{osdi22-alpa,
author = {Lianmin Zheng and Zhuohan Li and Hao Zhang and Yonghao Zhuang and Zhifeng Chen and Yanping Huang et al.},
title = {Alpa: Automating Inter and Intra-Operator Parallelism for Distributed Deep Learning},
booktitle = {16th USENIX Symposium on Operating Systems Design and Implementation (OSDI 22)},
year = {2022},
isbn = {978-1-939133-28-1},
address = {Carlsbad, CA},
pages = {559--578},

publisher = {USENIX Association},
month = jul
}

@inproceedings{osdi23-alpaserve,
author = {Zhuohan Li and Lianmin Zheng and Yinmin Zhong and Vincent Liu and Ying Sheng and Xin Jin and Yanping Huang et al.},
title = {AlpaServe: Statistical Multiplexing with Model Parallelism for Deep Learning Serving},
booktitle = {17th USENIX Symposium on Operating Systems Design and Implementation (OSDI 23)},
year = {2023},
isbn = {978-1-939133-34-2},
address = {Boston, MA},
pages = {663--679},

publisher = {USENIX Association},
month = jul
}

@misc{arxiv24-burstgp,
      title={BurstGPT: A Real-world Workload Dataset to Optimize LLM Serving Systems}, 
      author={Yuxin Wang and Yuhan Chen and Zeyu Li and Xueze Kang and Yuchu Fang and Yeju Zhou and Yang Zheng and Zhenheng Tang and Xin He and Rui Guo and Xin Wang and Qiang Wang and Amelie Chi Zhou and Xiaowen Chu},
      year={2025},
      eprint={2401.17644},
      archivePrefix={arXiv},
      primaryClass={cs.DC},
      url={https://arxiv.org/abs/2401.17644}, 
}

@article{arxiv19-megatron,
  author       = {Mohammad Shoeybi and Mostofa Patwary and Raul Puri and Patrick LeGresley and Jared Casper and Bryan Catanzaro},
  title        = {Megatron-LM: Training Multi-billion Parameter Language Models Using Model Parallelism},
  journal      = {arXiv preprint arXiv:1909.08053},
  year         = {2019}
}

@misc{23-alpaca,
  author       = {Rohan Taori and Ishaan Gulrajani and Tianyi Zhang and Yann Dubois and Xuechen Li and Carlos Guestrin and Percy Liang et al.},
  title        = {Stanford Alpaca: An Instruction-following LLaMA Model},
  year         = {2023},
  howpublished = {\url{https://github.com/tatsu-lab/stanford_alpaca}},
  note         = {GitHub repository, Accessed: 2025-05-04}
}

@misc{qin2024mooncakekvcachecentricdisaggregatedarchitecture,
      title={Mooncake: A KVCache-centric Disaggregated Architecture for LLM Serving}, 
      author={Ruoyu Qin and Zheming Li and Weiran He and Mingxing Zhang and Yongwei Wu and Weimin Zheng and Xinran Xu},
      year={2024},
      eprint={2407.00079},
      archivePrefix={arXiv},
      primaryClass={cs.DC},

}

@inproceedings{icsoc24-uellm,
author = {He, Yiyuan and Xu, Minxian and Wu, Jingfeng and Zheng, Wanyi and Ye, Kejiang and Xu, Chengzhong},
title = {UELLM: A Unified and Efficient Approach for Large Language Model Inference Serving},
year = {2024},
isbn = {978-981-96-0804-1},
publisher = {Springer-Verlag},
address = {Berlin, Heidelberg},
url = {https://doi.org/10.1007/978-981-96-0805-8_16},
doi = {10.1007/978-981-96-0805-8_16},
booktitle = {Service-Oriented Computing: 22nd International Conference, ICSOC 2024, Tunis, Tunisia, December 3–6, 2024, Proceedings, Part I},
pages = {218–235},
numpages = {18},
location = {Tunis, Tunisia}
}

@misc{ollama,
  title        = {Ollama - Get up and running with large language models.},
  howpublished = {\url{https://ollama.com/}},
  year         = {2025},
  note = {Accessed: Apr. 23, 2025},
}

@misc{gpt5.2,
  title        = {GPT-5.2},
  howpublished      = {https://openai.com/index/introducing-gpt-5-2/},
  year         = {2025},
  note = {Accessed: Dec. 20, 2025},
}

@article{DeepSeekV3TR,
  title={DeepSeek-V3 Technical Report},
  author={DeepSeek-AI and Aixin Liu and Bei Feng and Bing Xue and Bing-Li Wang et al.},
  journal={ArXiv},
  year={2024},
  volume={abs/2412.19437},

}

@article{LoongServe,
  title={LoongServe: Efficiently Serving Long-Context Large Language Models with Elastic Sequence Parallelism},
  author={Bingya Wu and Shengyu Liu and Yinmin Zhong and Peng Sun and Xuanzhe Liu and Xin Jin},
  journal={Proceedings of the ACM SIGOPS 30th Symposium on Operating Systems Principles},
  year={2024}
}

@inproceedings{choi2022serving,
  title={Serving heterogeneous machine learning models on Multi-GPU servers with Spatio-Temporal sharing},
  author={Choi, Seungbeom and Lee, Sunho and Kim, Yeonjae and Park, Jongse and Kwon, Youngjin and Huh, Jaehyuk},
  booktitle={2022 USENIX Annual Technical Conference (USENIX ATC 22)},
  pages={199--216},
  year={2022}
}

@Misc{xFormers2022,
  author =       {Benjamin Lefaudeux and Francisco Massa and Diana Liskovich and Wenhan Xiong and Vittorio Caggiano and Sean Naren and Min Xu and Jieru Hu and Marta Tintore and Susan Zhang and Patrick Labatut and Daniel Haziza and Luca Wehrstedt and Jeremy Reizenstein and Grigory Sizov},
  title =        {xFormers: A modular and hackable Transformer modelling library},
  howpublished = {\url{https://github.com/facebookresearch/xformers}},
  year =         {2022}
}

@inproceedings{wolf-etal-2020-transformers,
    title = "Transformers: State-of-the-Art Natural Language Processing",
    author = "Thomas Wolf and Lysandre Debut and Victor Sanh and Julien Chaumond and Clement Delangue and Anthony Moi and Pierric Cistac and Tim Rault and Rémi Louf and Morgan Funtowicz and Joe Davison and Sam Shleifer and Patrick von Platen and Clara Ma and Yacine Jernite and Julien Plu and Canwen Xu and Teven Le Scao and Sylvain Gugger and Mariama Drame and Quentin Lhoest and Alexander M. Rush",
    booktitle = "Proceedings of the 2020 Conference on Empirical Methods in Natural Language Processing: System Demonstrations",
    month = oct,
    year = "2020",
    address = "Online",
    publisher = "Association for Computational Linguistics",
    url = "https://www.aclweb.org/anthology/2020.emnlp-demos.6",
    pages = "38--45"
}

@article{Patel2023SplitwiseEG,
  title={Splitwise: Efficient Generative LLM Inference Using Phase Splitting},
  author={Pratyush Patel and Esha Choukse and Chaojie Zhang and Inigo Goiri and Aashaka Shah and Saeed Maleki and Ricardo Bianchini},
  journal={2024 ACM/IEEE 51st Annual International Symposium on Computer Architecture (ISCA)},
  year={2023},
  pages={118-132},

}

@misc{hu2024inferenceinterferencedisaggregatellm,
      title={Inference without Interference: Disaggregate LLM Inference for Mixed Downstream Workloads}, 
      author={Cunchen Hu and Heyang Huang and Liangliang Xu and Xusheng Chen and Jiang Xu and Shuang Chen and Hao Feng and Chenxi Wang and Sa Wang and Yungang Bao and Ninghui Sun and Yizhou Shan},
      year={2024},
      eprint={2401.11181},
      archivePrefix={arXiv},
      primaryClass={cs.DC},

}

@article{duan2024muxserve,
  title={MuxServe: Flexible Multiplexing for Efficient Multiple LLM Serving},
  author={Duan, Jiangfei and Lu, Runyu and Duanmu, Haojie and Li, Xiuhong and Zhang, Xingcheng and Lin, Dahua and Stoica, Ion and Zhang, Hao},
  journal={arXiv preprint arXiv:2404.02015},
  year={2024}
}

@article{qwen,
  title={Qwen3 technical report},
  author={Yang, An and Li, Anfeng and Yang, Baosong and Zhang, Beichen and Hui, Binyuan and Zheng, Bo and Yu, Bowen and Gao, Chang and Huang, Chengen and Lv, Chenxu and others},
  journal={arXiv preprint arXiv:2505.09388},
  year={2025}
}

@misc{Ali-cloud,
  title        = {Alibaba Cloud},
  author       = {Alibaba},
  howpublished = {Online},
  url          = {https://www.alibabacloud.com},
  note         = {Alibaba Cloud free service product catalog and introduction page. Accessed: 2026-01-23},
  year         = {2026},
  urldate      = {2026-01-23}
}

@misc{AzureML,
  author = {{Microsoft}},
  title = {Azure {Machine Learning}},
  howpublished = {\url{https://azure.microsoft.com/en-us/products/machine-learning}},
  year = {2024},
  note = {Accessed: 2024-05-20}
}

@misc{VertexAI,
  author = {{Google Cloud}},
  title = {Vertex {AI}},
  howpublished = {\url{https://cloud.google.com/vertex-ai}},
  year = {2024},
  note = {Accessed: 2024-05-20}
}

@misc{OpenRouter,
  author = {{OpenRouter}},
  title = {OpenRouter: {A} unified interface for {LLMs}},
  howpublished = {\url{https://openrouter.ai/}},
  year = {2024},
  note = {Accessed: 2024-05-20}
}

@misc{azure2024public,
  title = {Microsoft Azure Public Dataset},
  author = {{Microsoft Azure}},
  year = {2024},
  howpublished = {\url{https://github.com/Azure/AzurePublicDataset}},
  note = {Accessed: 2026-01-27}
}

@article{wan2025pipeoffload,
  title={Pipeoffload: Improving scalability of pipeline parallelism with memory optimization},
  author={Wan, Xinyi and Qi, Penghui and Huang, Guangxing and Lin, Min and Li, Jialin},
  journal={arXiv preprint arXiv:2503.01328},
  year={2025}
}

@inproceedings{Aegaeon,
  title={Aegaeon: Effective GPU pooling for concurrent LLM serving on the market},
  author={Xiang, Yuxing and Li, Xue and Qian, Kun and Yang, Yufan and Zhu, Diwen and Yu, Wenyuan and Zhai, Ennan and Liu, Xuanzhe and Jin, Xin and Zhou, Jingren},
  booktitle={Proceedings of the ACM SIGOPS 31st Symposium on Operating Systems Principles},
  pages={1030--1045},
  year={2025}
}

@inproceedings{ServerlessLLM,
title={ServerlessLLM: Low-Latency Serverless Inference for Large Language Models},
author={Fu, Yao and Xue, Leyang and Huang, Yeqi and Brabete, Andrei-Octavian and Ustiugov, Dmitrii and Patel, Yuvraj and Mai, Luo},
booktitle={18th USENIX Symposium on Operating Systems Design and Implementation (OSDI 24)},
pages={135--153},
year={2024}
}

@inproceedings{BlitzScale,
author = {Zhang, Dingyan and Wang, Haotian and Liu, Yang and Wei, Xingda and Shan, Yizhou and Chen, Rong and Chen, Haibo},
title = {BLITZSCALE: fast and live large model autoscaling with O(1) host caching},
year = {2025},
isbn = {978-1-939133-47-2},
publisher = {USENIX Association},
address = {USA},
booktitle = {Proceedings of the 19th USENIX Conference on Operating Systems Design and Implementation},
articleno = {16},
numpages = {19},
location = {Boston, MA, USA},
series = {OSDI '25}
}

@article{dean2008mapreduce,
  title={MapReduce: simplified data processing on large clusters},
  author={Dean, Jeffrey and Ghemawat, Sanjay},
  journal={Communications of the ACM},
  volume={51},
  number={1},
  pages={107--113},
  year={2008},
  publisher={ACM New York, NY, USA}
}

@inproceedings{zeng2025medusa,
  title={Medusa: Accelerating serverless LLM inference with materialization},
  author={Zeng, Shaoxun and Xie, Minhui and Gao, Shiwei and Chen, Youmin and Lu, Youyou},
  booktitle={Proceedings of the 30th ACM International Conference on Architectural Support for Programming Languages and Operating Systems, Volume 1},
  pages={653--668},
  year={2025}
}

@inproceedings{yang2022infless,
  title={Infless: a native serverless system for low-latency, high-throughput inference},
  author={Yang, Yanan and Zhao, Laiping and Li, Yiming and Zhang, Huanyu and Li, Jie and Zhao, Mingyang and Chen, Xingzhen and Li, Keqiu},
  booktitle={Proceedings of the 27th ACM International Conference on Architectural Support for Programming Languages and Operating Systems},
  pages={768--781},
  year={2022}
}

@article{lou2025hydraserve,
  title={HydraServe: Minimizing Cold Start Latency for Serverless LLM Serving in Public Clouds},
  author={Lou, Chiheng and Qi, Sheng and Jin, Chao and Nie, Dapeng and Yang, Haoran and Ding, Yu and Liu, Xuanzhe and Jin, Xin},
  journal={arXiv preprint arXiv:2502.15524},
  year={2025}
}

@article{lin2025flexpipe,
  title={FlexPipe: Adapting Dynamic LLM Serving Through Inflight Pipeline Refactoring in Fragmented Serverless Clusters},
  author={Lin, Yanying and Peng, Shijie and Lu, Chengzhi and Xu, Chengzhong and Ye, Kejiang},
  journal={arXiv preprint arXiv:2510.11938},
  year={2025}
}

@misc{nano-vllm,
  title        = {Nano-vLLM: A Lightweight vLLM Implementation Built from Scratch},
  author       = {GeeeekExplorer},
  year         = {2025},
  howpublished = {\url{https://github.com/GeeeekExplorer/nano-vllm}},
  note         = {GitHub repository},
}

\end{document}